\documentclass[12pt]{iopart}
\usepackage{iopams}  
\usepackage{hyperref}
\usepackage{bbm}
\usepackage{amsfonts, amssymb}
\usepackage{color}
\usepackage{graphicx,epsfig}
\usepackage{pifont} 
\usepackage{subfigure}
\begin{document}

\title{An all-optical feedback assisted steady state of an optomechanical array}

\author{Chaitanya Joshi}
\address{SUPA, Institute for Photonics and Quantum Sciences,
Heriot-Watt University, Edinburgh, EH14 4AS, UK}
\address{School of Physics and Astronomy, University of St Andrews, St  Andrews, KY16 9SS, UK }
\ead{cj30@st-andrews.ac.uk}
\author{Uzma Akram}
\address{Centre for Engineered Quantum Systems,
School of Mathematics and Physics,
The University of Queensland, St Lucia, QLD 4072, Australia}
\author{G. J. Milburn}
\address{Centre for Engineered Quantum Systems,
School of Mathematics and Physics,
The University of Queensland, St Lucia, QLD 4072, Australia}
\begin{abstract}
We explore the effect of all-optical feedback on the steady state dynamics of optomechanical arrays arising from various topologies. Firstly we consider an array comprised of a pair of independent optomechanical cavities coupled reversibly via their optical modes. Next we consider an optomechanical network formed from coupling two optical modes with interactions mediated via a common mechanical mode. Finally we extend the analysis to a large network of N-coupled optomechanical systems. Our results show implementing an-all optical feedback loop in each arrangement can enhance the degree of steady state entanglement between inter cavity optical and mechanical  modes.  
\end{abstract}
\pacs{42.50.Gy, 03.65.Ud, 03.67.Bg}
\maketitle
\noindent
\section{Introduction}
\label{sec:introdu}
The drive towards exploring quantum features at the mesoscopic and macroscopic scale has intensified in recent times with marked advances in understanding of the quantum-classical transition \cite{zure}. The prime motivations have been to test the limits of the quantum theory and to search for various quantum communication protocols \cite{orome} while discovering novel applications of mesoscopic systems as components of quantum memories and repeaters required in the implementation of various quantum information processing tasks \cite{Stannigel,Bagheri} and new sensor technology~\cite{mjwolley, gavartin}.  
 
 In this respect vastly different physical systems such as ultra-cold atoms coupled to nanomechanical systems \cite{ptre,Brennecke,murch,josh1}, superconducting qubits coupled to microwave resonators \cite{jamj}, opto- and electro- mechanical cavities \cite{florin,favero,jiang,wiederhecker,ma,park,kippenbergreview,APS-Physics,isart1,isart2,isart3, chang, regal, corbitt,sbos, Schliesser, eichenfield, Thompson, cleland, markusstrongcoupling,vacoptomechrate, lahaye,naik,teufel,josh2} have recently attracted a lot of attention in probing quantum signatures at the mesoscopic scale. By and large the goal has been to achieve coherent control over the motion of the mechanical element via coupling to electromagnetic and optical modes in the quantum regime \cite{aspelmeyerreview}. Such a coupling inevitably has the potential to engineer a hybrid quantum device composed of otherwise incompatible degrees of freedom of different physical systems. Among the plethora of such proposals, special attention has been devoted to explore further the potentials of optomechanics, in particular. Optomechanics makes use of radiation pressure of light on a mechanical element resulting in reversible coupling between optics and mechanics, two vastly different physical systems at the extremes of the quantum world. Recent progress in the field not only promises to strongly couple these very disparate modes identifying a whole new class of engineered quantum systems but also to control the dynamics of such systems, \cite{asp_kippen_marquardt}. 
 
 In an optomechanical (OM) set-up, the resonance frequency of the optical cavity under consideration can be altered by the displacement of the mechanical element which may for example form one of the mirrors in a Fabry-P\'{e}rot cavity. As a result of radiation pressure of light, reversible coupling is generated between the optical and  mechanical modes. The strength of the OM coupling is however derived from the magnitude of the radiation pressure force which can be enhanced by strong coherent driving in the considered OM system. As has been extensively shown in the resolved sideband regime, \cite{vitali, muller, hartmann, vitali2,paternostro2,genes,tiaan1,tiaan}  OM interaction can generate entanglement between the optical and mechanical modes within an OM system. Further for two or more coupled OM systems forming an OM array, intra cavity OM entanglement can be distributed over inter cavity modes in the steady state \cite{josh2,paternostro,akram}. In the present work we are interested in devising a scheme whereby entanglement can not only be generated but can also be {\em controlled} as well. The ability to control entanglement would be a useful resource in the design and implementation of quantum repeaters and quantum memories crucial in the implementation of quantum information processing. It may also enable new measurement protocols based on mechanical sensors, for example, weak force sensors~\cite{gavartin, houxun}. With this objective in mind we apply an all-optical feedback scheme to earlier proposals, \cite{akram} considering coupled optomechanical arrays. All-optical feedback was considered extensively in the context of cascaded cavities \cite{wiseman} and associated amplitude and quadrature feedback. 

We apply similar coupling configurations to a system of OM cavities in our analysis and compare the steady state dynamics of various OM networks both in the presence and absence of an all-optical feedback loop. Specifically we address two different topologies: Firstly we consider a pair of independent but reversibly coupled OM systems. Secondly we consider an OM arrangement where two different optical modes are coupled via a common mechanical element. In fact our results also apply to a hybrid system in which one optical mode and one microwave mode interact via a common mechanical element as in~\cite{andrews, barzanjeh}. In each case the all-optical feedback loop is implemented by an additional irreversible coupling between the two optical modes. Our results show that while a reversibly coupled OM network can result in steady state entanglement between the inter cavity mechanical modes, implementing an all optical feedback loop can enhance the generated entanglement. Hence an all-optical feedback can be used as a resource to control entanglement in the steady state of an OM network. Furthermore in the last section of this work, we extend our analysis to a large array comprising of reversibly coupled OM ports. Each OM port in the array is composed of a pair of OM systems coupled via an all-optical feedback loop. Our results show bipartite steady state entanglement only exists between mechanical resonators from different OM ports with different parity, and can be enhanced in the presence of feedback.

 \section{Feedback in an optomechanical network} 
 \label{sec:mod} 
\begin{figure}[!h]
\centering
\includegraphics[scale=0.8]{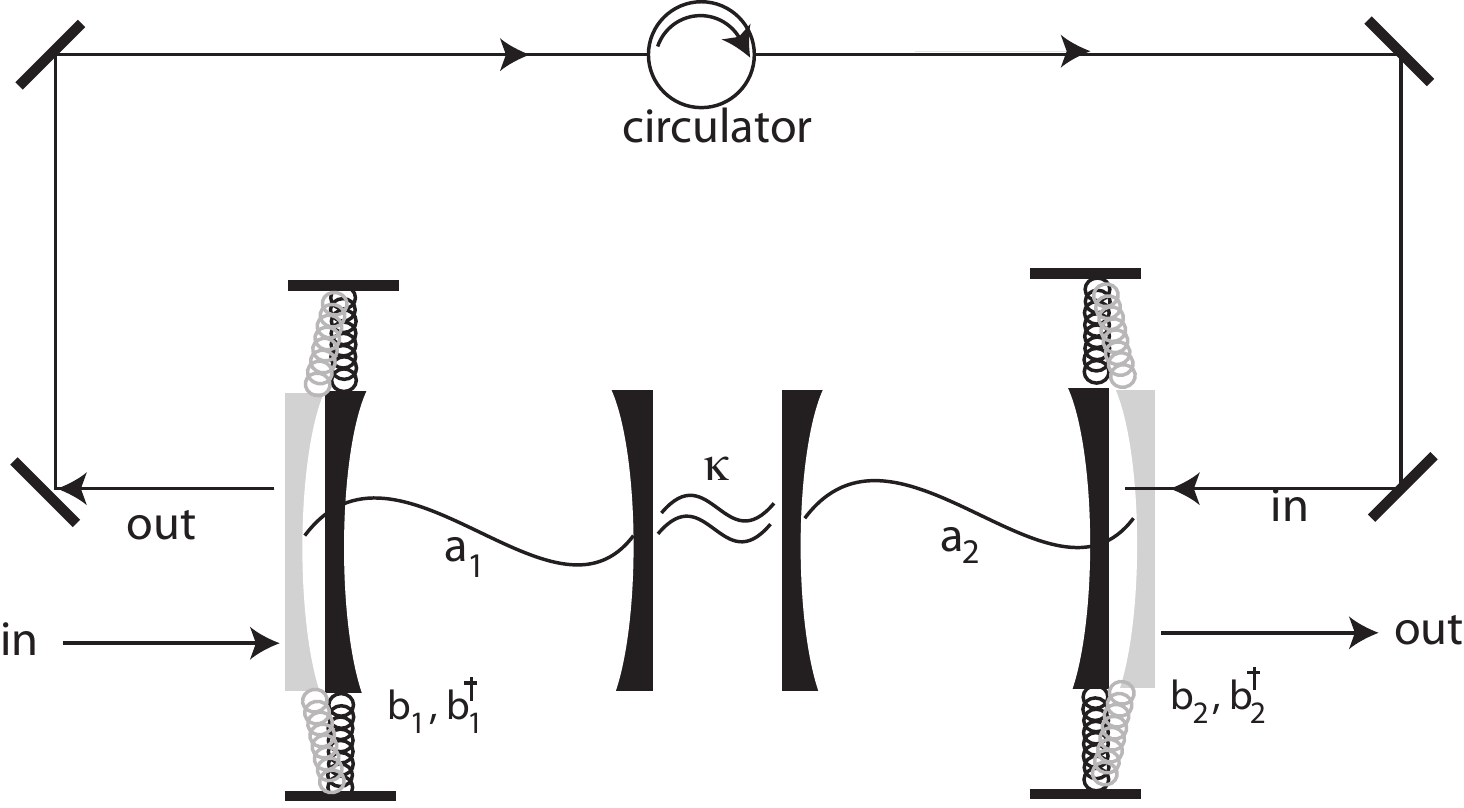}
\caption{ Model 1: Implementing an all-optical feedback scheme in an OM array: Two OM cavities are coupled reversibly with a coupling strength $\kappa$ via neighboring optical modes $a_{1,2}$ as well as irreversibly with mode $\hat{a}_{1}$ driving $\hat{a}_{2}$ in a forward feed direction. The optical mode $\hat{a}_{j}$ and mechanical mode $\hat{b}_{j}$ in each cavity are coupled reversibly by an OM coupling strength $g_{j}$.}. 
\label{model_setup1}
\end{figure}
\subsection{Two coupled optomechanical cavities}
\label{model_1}
We begin by first considering the set-up in Fig.~\ref{model_setup1} which comprises of two OM cavities, (OM 1 and 2) coupled reversibly via their optical modes. Additionally the optical mode $a_{1}$ of OM 1 is also coupled irreversibly to  mode $a_{2}$ of OM 2. Simultaneous presence of unidirectional and bidirectional  coupling between the optical modes constitute an all-optical feedback loop \cite{wiseman}. The system OM 1 acts as a source driving the state of the OM 2 cavity which in turn influences the state of source cavity through reversible coupling of arbitrary strength $\kappa$.

 In the usual convention, radiation pressure is induced by strong external coherent driving on each OM system hence displacing the cavity field by its steady state amplitude. The resulting OM interaction is then linearized such that a quadratic  coupling is induced between the optical and mechanical modes in each system \cite{akram}. In this linearized regime and in the rotating frame of the external drive,  the OM interaction between each individual cavity takes the form (in units $\hbar=1$): \cite{akram}
 \begin{eqnarray}
{H}_{I}&=& \sum_{j=1}^{2} \Delta_{j}\hat{a}_{j}^{\dagger}\hat{a}_{j}+\omega_{m_j}\hat{b}_{j}^{\dagger}\hat{b}_{j}+ g_{j}(\hat{a}_{j}+ \hat{a}_{j}^{\dagger})(\hat{b}_{j}+ \hat{b}_{j}^{\dagger}),
\label{HI_modelfull}
\end{eqnarray}
 where $\Delta_{j}=\Omega_{j}-\omega_{L_{j}}$, where $\omega_{L_{j}}$ is the frequency of the external laser drive pumping  the $j$-th cavity with resonance frequency $\Omega_{j}$, $\omega_{m_j}$ is the fundamental resonance frequency of the $j$-th mechanical mode, $g_{j}$ is the effective optomechanical coupling strength proportional to the steady state amplitude of the cavity field due to linearization of the radiation pressure force, and $\hat{a}_{j}(\hat{b}_{j})$ and $\hat{a}_{j}^{\dagger}(\hat{b}_{j}^{\dagger})$  are the respective annihilation and creation operators for the optical (mechanical) modes for each cavity. In the presence of the OM interaction of the form \eref{HI_modelfull}, the optical and mechanical modes of each individual OM network will become quantum entangled. In the present work we are interested in achieving long-distance entanglement in coupled OM cavity arrays. Moreover, we want to explore the influence of an all-optical feedback on inter-cavity entanglement. In the presence of an all-optical feedback and under the Born-Markov approximation \cite{milburn}, the open system dynamics of the coupled OM cavities is given by the following master equation:
 \begin{eqnarray}
\label{ME_modelnew1}
\frac{d\rho}{dt}&=& -i [{H}_I,\rho]-i\kappa \left[\hat{a}_1 \hat{a}_2^\dagger e^{i(\omega_{L_2} -\omega_{L_1})t}+\hat{a}_2 \hat{a}_1^\dagger e^{i(\omega_{L_1} -\omega_{L_2})t}, \rho\right] \nonumber \\&&+\sqrt{\Gamma_{1}\Gamma_{2}}([\hat{a}_{1}\rho,\hat{a}_{2}^{\dagger}]e^{i(\omega_{L_2} -\omega_{L_1})t}+[\hat{a}_{2},\rho \hat{a}_{1}^{\dagger}]e^{i(\omega_{L_1} -\omega_{L_2})t}) \nonumber \\
&& +\sum_{j=1}^2\Gamma_j{\cal D}[\hat{a}_j]\rho +\gamma_{j} (\bar{n}_{j}+1){\cal D}[\hat{b}_{j}]\rho+\gamma_{j} \bar{n}_{j}{\cal D}[\hat{b}_{j}^\dagger]\rho,
\end{eqnarray}
where the linewidth of each OM cavity is given by $\Gamma_{j}$ while the mechanical resonators are allowed to decay to a thermal bath with average thermal occupancy $\bar{n}_{j}$ at a rate $\gamma_{j}$. 
The damping superoperator in equation \eref{ME_modelnew1} ${\cal D}[A]$ is defined in the usual way by
\begin{equation}
{\cal D}[A]\rho=A\rho A^\dagger -\frac{1}{2} (A^\dagger A\rho+\rho A^\dagger A).
\end{equation}

In the interaction picture of the unitary operator $\hat{U}=e^{\sum_{j=1}^{2}i(\Delta_{j}\hat{a}^{\dagger}_{j}\hat{a}_{j}+\omega_{m_j}\hat{b}^{\dagger}_{j}\hat{b}_{j})t}$, the master equation \eref{ME_modelnew1} takes the form:
 \begin{eqnarray}
\label{ME_modelnew2}
\frac{d\rho}{dt}&=& -i \sum_{j=1}^{2}g_{j}[(\hat{a}_{j}e^{-i \Delta_{j} t}+\hat{a}^{\dagger}_{j}e^{i \Delta_{j} t})(\hat{b}_{j}e^{-i \omega_{m_j} t}+\hat{b}^{\dagger}_{j}e^{i \omega_{m_j}t}),\rho]\nonumber \\
&&-i\kappa \left[\hat{a}_1 \hat{a}_2^\dagger e^{i(\omega_{L_2} -\omega_{L_1}+ \Delta_{2}- \Delta_{1})t}+\hat{a}_2 \hat{a}_1^\dagger e^{i(\omega_{L_1} -\omega_{L_2}+ \Delta_{1}- \Delta_{2})t}, \rho\right] \nonumber \\&&+\sqrt{\Gamma_{1}\Gamma_{2}}([\hat{a}_{1}\rho,\hat{a}_{2}^{\dagger}]e^{i(\omega_{L_2} -\omega_{L_1}+ \Delta_{2}- \Delta_{1})t}+[\hat{a}_{2},\rho \hat{a}_{1}^{\dagger}]e^{i(\omega_{L_1} -\omega_{L_2}+ \Delta_{1}- \Delta_{2})t}) \nonumber \\
&& +\sum_{j=1}^2\Gamma_j{\cal D}[\hat{a}_j]\rho +\gamma (\bar{n}+1){\cal D}[\hat{b}_{j}]\rho+\gamma \bar{n}{\cal D}[\hat{b}_{j}^\dagger]\rho. 
\end{eqnarray}

 For simplicity, we take both OM systems to be identical such that $\Omega_{1}=\Omega_{2}$, both mechanical resonators have the same frequency, $\omega_{m_{j}}=\omega_{m}$, mechanical damping rate $\gamma_{j}=\gamma$ and thermal occupation number $\bar{n}_{j}=\bar{n}$. Each OM cavity is however driven on a different sideband. In particular we drive OM 1 exclusively on the blue sideband, $\Delta_{1}=-\omega_{m}$ and OM 2 on the red sideband, $\Delta_{2}=\omega_{m}$. Under the rotating-wave approximation (RWA) \cite{akram,milburn}, the OM interaction simplifies to the following form 
\begin{eqnarray}
{\tilde{H}}_{I}&=& g_{1}(\hat{a}_{1}\hat{b}_{1} + \hat{a}_{1}^{\dagger}\hat{b}_{1}^{\dagger}) +  g_{2}(\hat{a}_{2}^{\dagger}\hat{b}_{2}+\hat{a}_{2}\hat{b}_{2}^{\dagger}),
\label{HI_model1}
\end{eqnarray}
and the master equation describing the open system dynamics of the coupled OM network takes the following form
\begin{eqnarray}
\label{ME_model1}
\frac{d\rho}{dt}&=& -i [{\tilde{H}}_I,\rho]-i\kappa \left[\hat{a}_1 \hat{a}_2^\dagger  +\hat{a}_2 \hat{a}_1^\dagger, \rho\right] +\sqrt{\Gamma_{1}\Gamma_{2}}([\hat{a}_{1}\rho,\hat{a}_{2}^{\dagger}]+[\hat{a}_{2},\rho \hat{a}_{1}^{\dagger}]) \nonumber \\
&& +\sum_{j=1}^2\Gamma_j{\cal D}[\hat{a}_j]\rho +\gamma (\bar{n}+1){\cal D}[\hat{b}_{j}]\rho+\gamma \bar{n}{\cal D}[\hat{b}_{j}^\dagger]\rho. 
\end{eqnarray}
The first two commutators  on the right hand side of the above equation denotes the unitary evolution of the coupled OM network, the successive two commutators represents the irreversible coupling between the optical modes of the first and the second cavity and the last six terms takes into account the damping of optical and mechanical modes. Reversible coupling between the optical modes can be generated either through evanescent coupling of the adjacent optical modes or through an optical fiber \cite{josh2}. Irreversible coupling between the source and the driven cavity can be established using an optical circulator \cite{akram,akram2}.  A non-reciprocal optical device such as Faraday rotator can be used to establish irreversible coupling between the optical modes of two adjacent OM cavities. 

This particular choice of detuning configurations in such a coupled array induces a two mode squeezing interaction between the optical and mechanical modes in OM 1, and a beam splitter interaction resulting in state transfer between optical and mechanical modes in OM 2. This configuration of coupled dynamics has been analyzed in detail in \cite{akram}, where two independent coupling configurations between individual OM systems were analyzed, i.e. the OM systems were considered to be {\em either} reversibly {\em or} irreversibly coupled via their optical ports. In the present work, however we extend the consideration to an {\em added} coupling channel between coupled OM systems 1 and 2. This additional coupling is engineered as an unidirectional coupling emanating from the optical mode of OM cavity 1 to the optical mode of OM cavity 2, while {\em also} maintaining a reversible coupling between the two inter cavity optical modes:  hence implementing an all-optical feedback loop, \cite{wiseman} in the optomechanical array. As has been established previously, coupling between such driven OM systems results in distribution of entanglement generated between optical and mechanical modes in OM cavity 1 over inter cavity modes in the steady state, \cite{akram}. However in the present work we are interested in exploring the possibility of {\em controlling/altering} the steady state inter cavity phonon-phonon entanglement via an all-optical feedback loop. 

Alternatively such a network of coupled cascaded quantum systems shown in Fig.\ref{model_setup1} can also be conveniently modeled using the SLH framework, \cite{SLH1,SLH2}. The SLH formalism is useful in forming reducible networks of coupled open quantum systems with the objective of implementing quantum control analysis and design. In the SLH formalism, an open quantum system is described by a set of three parameters, $G=(S,L,H)$ where $S$ denotes the scattering matrix, $L$ is the coupling vector and $H$ is the Hamiltonian operator for the system. Hence the set of coupled quantum OM systems summarized in Fig.~\ref{model_setup1} forms a series product connection under the SLH formalism \cite{SLH1}, 
\begin{eqnarray}
{\bf G_{2}} \triangleleft {\bf G_{1}}&=&({\bf S}_{\rm eff},{\bf L}_{\rm eff},{\bf H_{\rm eff}}) \\ \nonumber 
&=&\left({\bf S_{2}}{\bf S_{1}},{\bf L_{2}}+{\bf S_{2}}{\bf L_{1}},H_{1}+H_{2}+\frac{1}{2i}({\bf L_{2}}^{\dagger}{\bf S_{2}}{\bf L_{1}}-{\bf L_{1}}^{\dagger}{\bf S_{2}}^{\dagger}{\bf L_{2}})\right).
\end{eqnarray}
where the specific variables of the series product are ${\bf S_{j}}={\bf I}$, ${\bf L_{j}}=\sqrt{\Gamma_{j}}\hat{a}_{j}$ corresponding to each OM system and the combined coupling vector is ${\bf L}_{\rm eff}=\sqrt{\Gamma_{1}}\hat{a}_{1} +\sqrt{\Gamma_{2}}\hat{a}_{2}$. The Hamiltonian operator of the combined system is given as,
\begin{eqnarray}
{\bf H_{eff}}={\tilde{H}}_{I}+\kappa(\hat{a}_1 \hat{a}_2^\dagger +\hat{a}_2 \hat{a}_1^\dagger)+\frac{1}{2i}\left({\bf L_{2}}^{\dagger}{\bf S_{2}}{\bf L_{1}}- {\bf L_{1}^{\dagger}}{\bf S_{2}}{\bf L_{2}}\right)
\end{eqnarray} 
where ${\tilde{H}}_{I}$ has been defined earlier in equation \eref{HI_model1}. The Master equation of the reduced network can then be deduced from the three variables as \cite{SLH1},
\begin{eqnarray}
\frac{d}{dt}\rho=i[\rho,{\bf H_{eff}}]+{\cal D}[{\bf L_{eff}}]\rho. 
\label{ME_SLH}
\end{eqnarray}
Substituting the  expressions for ${\bf H_{eff}}$ and ${\cal D}[{\bf L_{eff}}]$ in the above equation one gets:
\begin{eqnarray}
 \frac{d}{dt}\rho &=&-i [{\tilde{H}}_{I},\rho]-i\kappa [\hat{a}_1 \hat{a}_2^\dagger +\hat{a}_2 \hat{a}_1^\dagger, \rho]+\Gamma_{1}\mathcal D[\hat{a}_{1}]\rho +\Gamma_{2} \mathcal D[\hat{a}_{2}]\rho \nonumber \\
 &&-\frac{\sqrt{\Gamma_{1}\Gamma_{2}}}{2}[\hat{a}_1 \hat{a}_2^\dagger -\hat{a}_2 \hat{a}_1^\dagger,\rho ]+ \sqrt{\Gamma_{1}\Gamma_{2}}(\hat{a}_{1}\rho \hat{a}_{2}^{\dagger}-\frac{1}{2}\hat{a}_{1}^{\dagger}\hat{a}_{2}\rho-\frac{1}{2}\rho\hat{a}_{1}^{\dagger}\hat{a}_{2})\nonumber \\
 &&+\sqrt{\Gamma_{1}\Gamma_{2}}(\hat{a}_{2}\rho \hat{a}_{1}^{\dagger}-\frac{1}{2}\hat{a}_{2}^{\dagger}\hat{a}_{1}\rho-\frac{1}{2}\rho\hat{a}_{2}^{\dagger}\hat{a}_{1}).\nonumber
\label{ME_SLH1}
\end{eqnarray}
Simplifying the above equation we get
\begin{eqnarray}
 \frac{d}{dt}\rho &=&-i [{\tilde{H}}_{I},\rho]-i\kappa [\hat{a}_1 \hat{a}_2^\dagger +\hat{a}_2 \hat{a}_1^\dagger, \rho]+\Gamma_{1}\mathcal D[\hat{a}_{1}]\rho +\Gamma_{2} \mathcal D[\hat{a}_{2}]\rho  \nonumber \\
 &&+ \sqrt{\Gamma_{1}\Gamma_{2}}(\hat{a}_{1}\rho \hat{a}_{2}^{\dagger}+\hat{a}_{2}\rho \hat{a}_{1}^{\dagger}-\hat{a}_{1} \hat{a}_{2}^{\dagger} \rho-\rho\hat{a}_{2} \hat{a}_{1}^{\dagger} ).
\label{ME_SLH2}
\end{eqnarray}
 Taking into account the damping of mechanical modes $b_{1}$ and $b_{2}$ by coupling them to identical independent thermal reservoirs, the dynamics of the coupled OM network is then given by 
 \begin{eqnarray}
 \frac{d}{dt}\rho &=&-i [{\tilde{H}}_{I},\rho]-i\kappa [\hat{a}_1 \hat{a}_2^\dagger +\hat{a}_2 \hat{a}_1^\dagger, \rho] + \sqrt{\Gamma_{1}\Gamma_{2}}(\hat{a}_{1}\rho \hat{a}_{2}^{\dagger}+\hat{a}_{2}\rho \hat{a}_{1}^{\dagger}-\hat{a}_{1} \hat{a}_{2}^{\dagger} \rho-\rho\hat{a}_{2} \hat{a}_{1}^{\dagger} )\nonumber \\
 &&+\sum_{j=1}^2\Gamma_j{\cal D}[\hat{a}_j]\rho +\gamma (\bar{n}+1){\cal D}[\hat{b}_{j}]\rho+\gamma \bar{n}{\cal D}[\hat{b}_{j}^\dagger]\rho.
\label{ME_SLH3}
\end{eqnarray}
It is thus clear  that the master equation \eref{ME_SLH3} is identical to the master equation \eref{ME_model1} of the OM network.

It  is easy to see from equation \eref{ME_model1} that the Hamiltonian, the Lindblad operators and the all-optical feedback loop are all bilinear in bosonic operators. Thus it is guaranteed that if the coupled OM array is initially prepared in a Gaussian state, then the evolution described by the Master equation \eref{ME_model1} maintains this Gaussian character. Thus to solve the Master equation \eref{ME_model1} for an initial Gaussian state, it is consistent to make a Gaussian ansatz for the normal-ordered quantum characteristic function of the following form  $\chi(\epsilon,\eta,x,y,{\it t}) =\exp\left[-z^{T}{\mbox{\bf A}}(t)\,z+i z^{T}h(t)\right]$, where ${\mbox{\bf A}}(t)$ is a time-dependent 8$\times$8 matrix and $h(t)$ is a 1$\times$8 time-dependent vector and  $z^{T}=(\epsilon,\epsilon^{*},\eta,\eta^{*},x,x^{*},y,y^{*})$. An explicit solution of the Master equation \eref{ME_model1} is outlined in the \ref{sec:appendixone}. From the  solution of the normal-ordered  quantum characteristic function it is a straightforward exercise to calculate all the correlators between the desired modes. 

To quantize the entanglement between any two modes, we utilize the logarithmic negativity as a measure of entanglement  for Gaussian states, \cite{gera}. For a two-mode Gaussian continuous variable state characterized by its covariance matrix $\bf V$, the logarithmic negativity  is computed  as 
$ \mathcal N= \rm{Max}[0,-\rm{log}(2 \nu_{-})]$ \cite{gera}
 where $\nu_{-}$ is the smallest of the symplectic eigenvalues of the covariance matrix given by  $
 \nu_{-}= \sqrt{\sigma/2-\sqrt{(\sigma^{2}-4 \rm{Det} \bf{V})}/\rm{2}}$. Here
$ \sigma=\rm{Det} \bf{A_{1}}+\rm{Det} \bf{B_{1}}-\rm{2Det} \bf{C_{1}}$ and
$\bf V=
\left(
\begin{array}{cc}
   \bf A_{1}  &\bf C_{1}\\
  \bf C_{1}^{T}  & \bf B_{1}\,   \\
\end{array}
\right)$,
 where $ \bf A_{1}$ ($\bf B_{1}$) accounts for the local variances of the two modes 
 and $\bf C_{1} $ for the inter-mode
correlations. 

\begin{figure}[ht!]
     \begin{center}
    \subfigure[$\bar{n}=0$]{%
           \label{fig:firstgeomb}
           \includegraphics[width=0.45\textwidth]{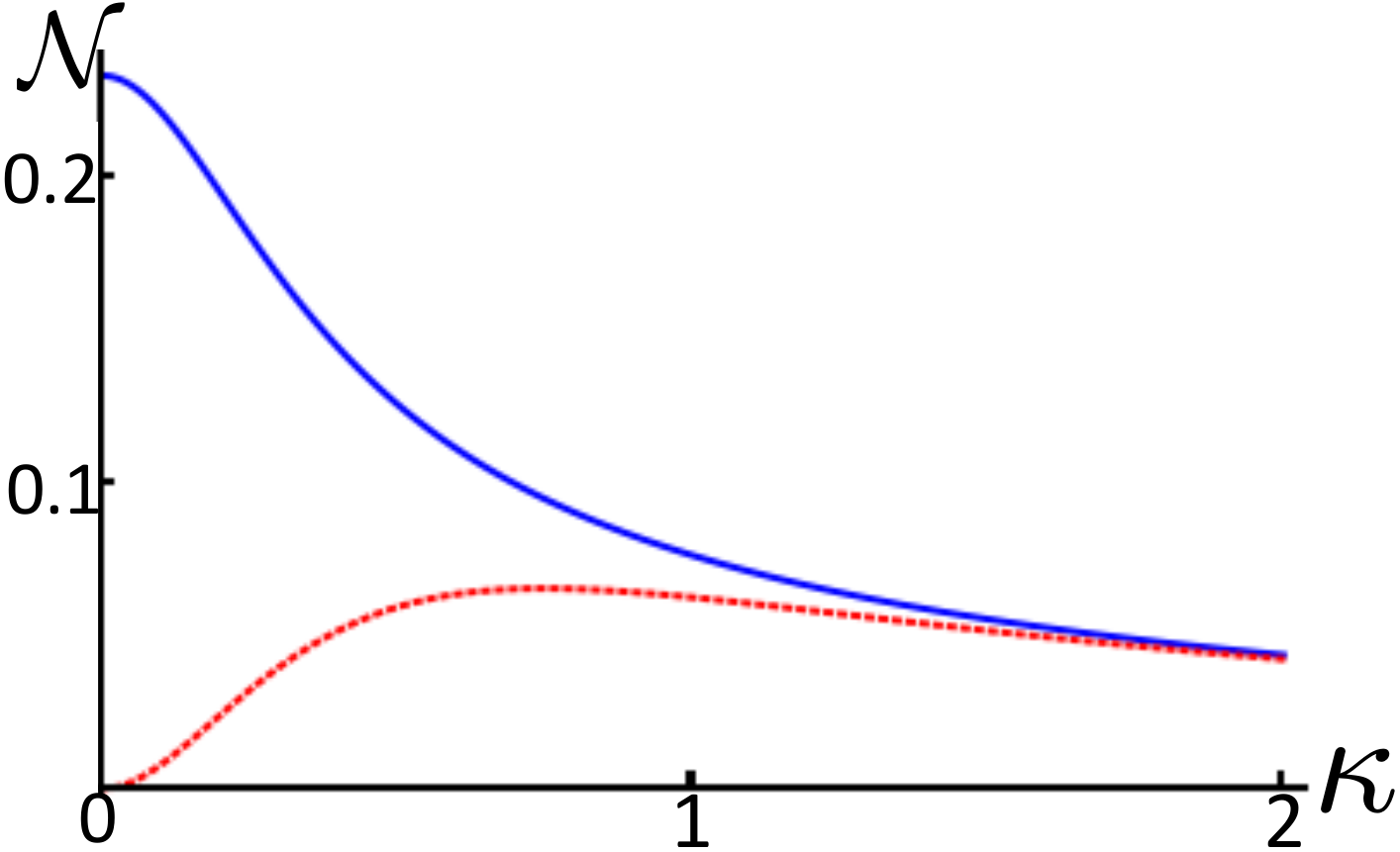}
        }
          \subfigure[$\kappa=0.1$]{%
            \label{fig:firstgeoma}
            \includegraphics[width=0.45\textwidth]{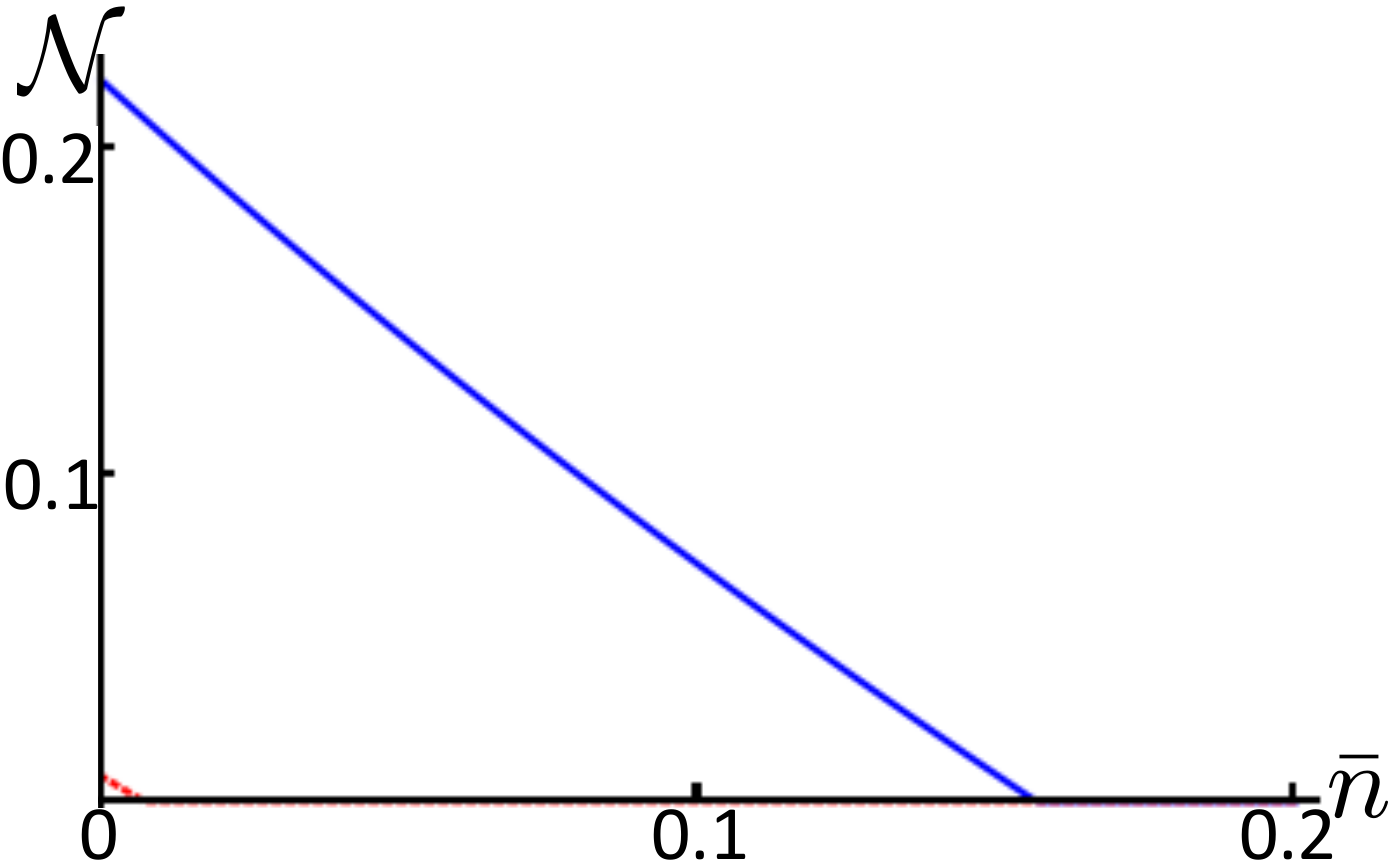}
        }
    \end{center}
    \caption{(Color online) Steady state entanglement between the two mechanical modes in presence (solid) and absence (dashed) of  an all-optical feedback. Physical parameters (in units of $\Gamma_{1}=\Gamma_{2}=\Gamma$): $\gamma_{1}=\gamma_{2}=10^{-2}, g_{1}=0.01, g_{2}=0.05$.
     }
   \label{fig:subfigures}
\end{figure}

We compute logarithmic negativity measuring the quantum entanglement between the mechanical modes $b_{1}$ and $b_{2}$ in the steady state and in Fig.~\ref{fig:firstgeomb} and Fig.~\ref{fig:firstgeoma} plot it as a function of the reversible coupling strength between the optical modes $\kappa$ and the thermal occupancy of each mechanical mode $\bar{n}$. We find finite steady state entanglement between the two inter cavity mechanical modes in the steady state. We observe the entanglement persists for a finite amount of nonzero thermal occupation number $\bar{n}$. We also note the entanglement between the mechanical modes is higher for smaller values of coupling strength  $\kappa$ between the optical modes.

\begin{figure}[ht!]
     \begin{center}

  \subfigure[]{%
            \label{fig:firstgeomc}
            \includegraphics[width=0.45\textwidth]{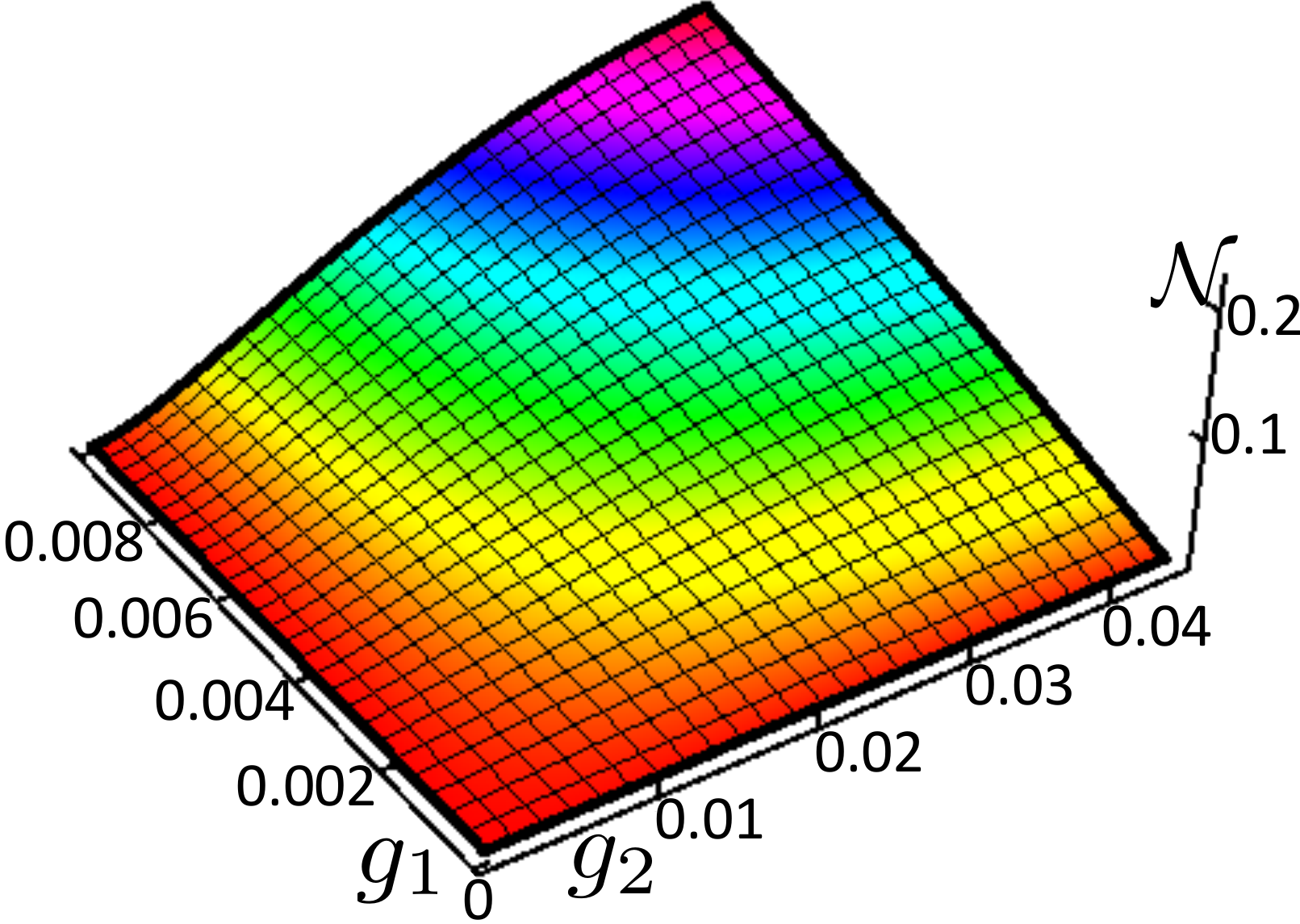}
        }
    \subfigure[]{%
           \label{fig:firstgeomd}
           \includegraphics[width=0.45\textwidth]{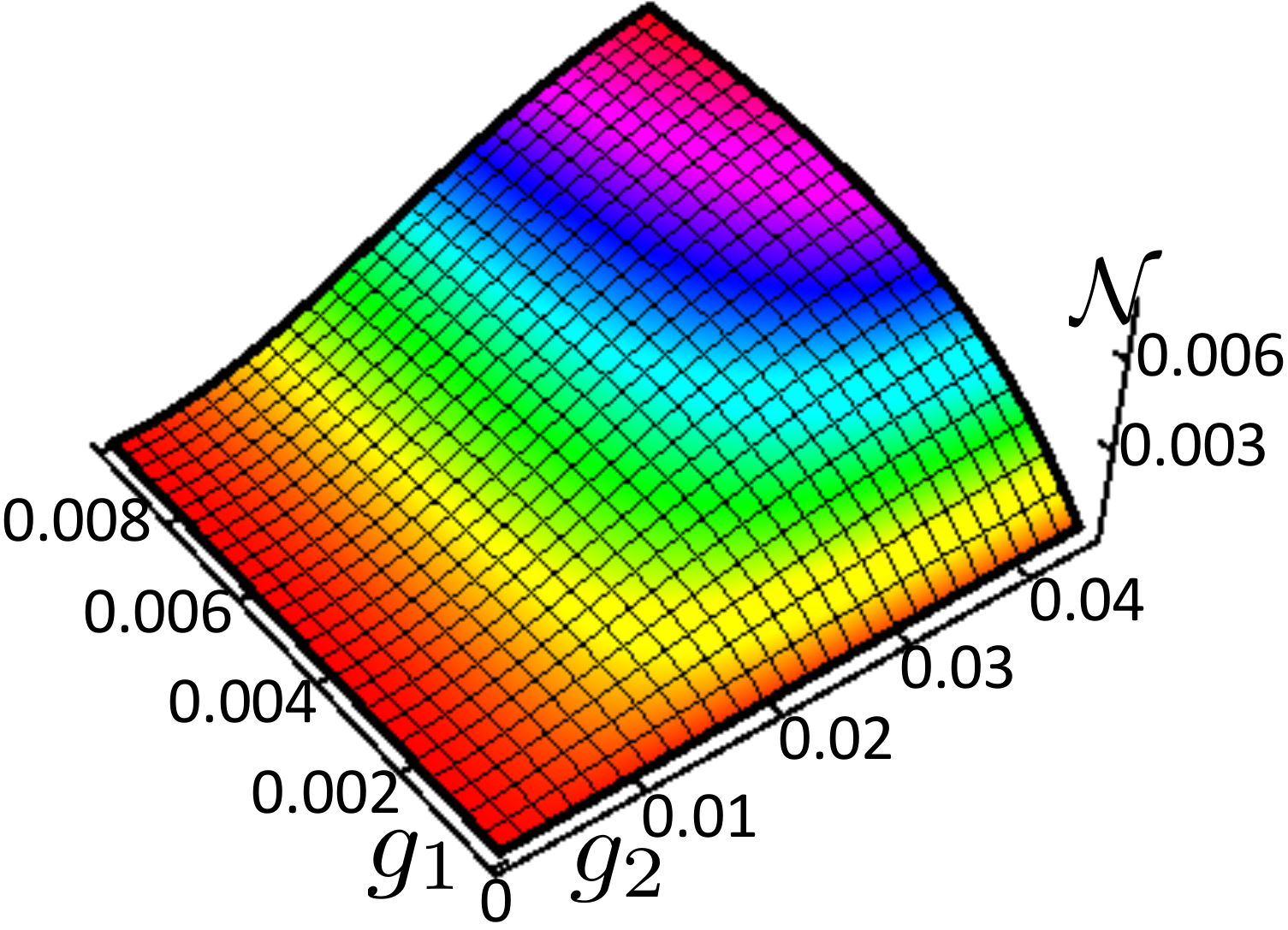}
        }
    \end{center}
    \caption{(Color online) Steady state entanglement between the two mechanical modes plotted as a function of $g_{1}$ and $g_{2}$ in (a) presence and (b) absence of  an all-optical feedback. Physical parameters (in units of $\Gamma_{1}=\Gamma_{2}=\Gamma$): $\gamma_{1}=\gamma_{2}=10^{-2}, \kappa=0.1$ and $\bar{n}=0$.
     }%
   \label{fig:subfigures1}
\end{figure}
To explore the advantage of using an all-optical feedback, in Fig.~\ref{fig:firstgeomb} and Fig.~\ref{fig:firstgeoma} we also plot the entanglement  between the mechanical modes $b_{1}$ and $b_{2}$ in absence of all-optical  feedback. Clearly the presence of feedback increases the degree of entanglement between inter cavity phonons. It can be seen from Fig.~\ref{fig:firstgeomb} that presence of reversible and irreversible couplings between the optical modes does have an impact on the entanglement between the mechanical modes. The influence is clearly marked for small values of $\kappa$ and under which case feedback generates stronger entanglement between the mechanical modes. From Fig.~\ref{fig:firstgeoma} it is clear that an all-optical feedback results in generating entangled state of mechanical modes which is more robust to thermal fluctuations. 

It is also interesting to see how the inter cavity phonon-phonon entanglement $\mathcal{N}$, varies as a function of the different optomechanical coupling strengths $g_{1}$ and $g_{2}$  in the OM array. These are plotted in Figs.~\ref{fig:firstgeomc} and Figs.~\ref{fig:firstgeomd} in the presence and absence of feedback respectively. Hence the added coupling channel between the inter cavity optical modes forming the feedback loop, can be used to increase the steady state entanglement in such an OM array.  

\begin{figure}[ht!]
     \begin{center}
       {
            \includegraphics[width=0.45\textwidth]{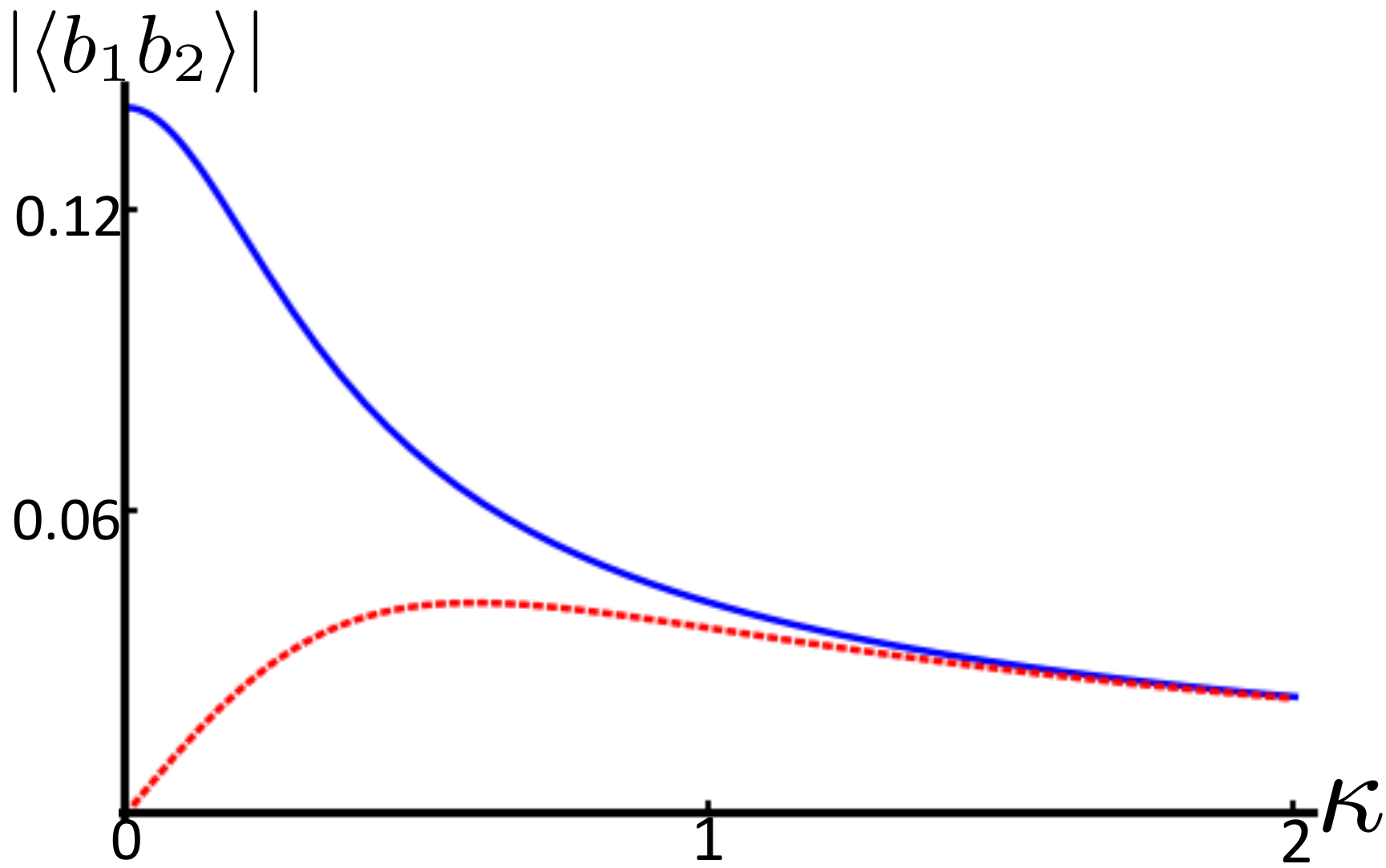}
        }
    \end{center}
    \caption{%
      (Color online) Absolute value of  steady state correlator $\langle \hat{b}_{1}\hat{b}_{2}\rangle$ plotted as a function of $\kappa$ in presence (solid) and absence (dashed) of an all-optical feedback.  Physical parameters (in units of $\Gamma_{1}=\Gamma_{2}=\Gamma$): $g_{1}=0.01, g_{2}=0.05, \gamma_{1}=\gamma_{2}=10^{-2}, \bar{n}=0$.
     }\label{fig:appndxfig}
\end{figure}

  To gain some more insight into the constructive role played by an all-optical  feedback in coupled OM array we consider a pragmatic regime where the linewidth of each cavity mode is very large compared to all other system parameters. Under this regime both the cavity modes can be adiabatically eliminated and stochastic differential equations for the mechanical modes can be arrived at. The detailed calculation is outlined in the \ref{sec:appndxthr}. The effect of feedback is clearly imprinted on the steady state correlators between the mechanical modes given by eqns. \eref{fedbckcorr} and \eref{nofedbckcorr}. The steady state correlator $|\langle \hat{b}_{1} \hat{b}_{2} \rangle|$ in presence and absence of feedback is plotted in Fig.~\ref{fig:appndxfig}. As can be clearly seen from Fig.~\ref{fig:appndxfig}, feedback has a clear influence on quantum correlations between the two mechanical modes. In particular for low values of $\kappa$, feedback helps in building stronger quantum correlations between the two mechanical modes. For values of $\kappa \sim \Gamma$, there is no evident difference in the strength of quantum correlations between the mechanical modes in the presence and absence of all-optical feedback in OM network array. This observation is also imprinted in Fig.~\ref{fig:firstgeomb}.

\subsection{Mechanics assisted optical coupling} 
\begin{figure}[!h]
\centering
\includegraphics[scale=0.8]{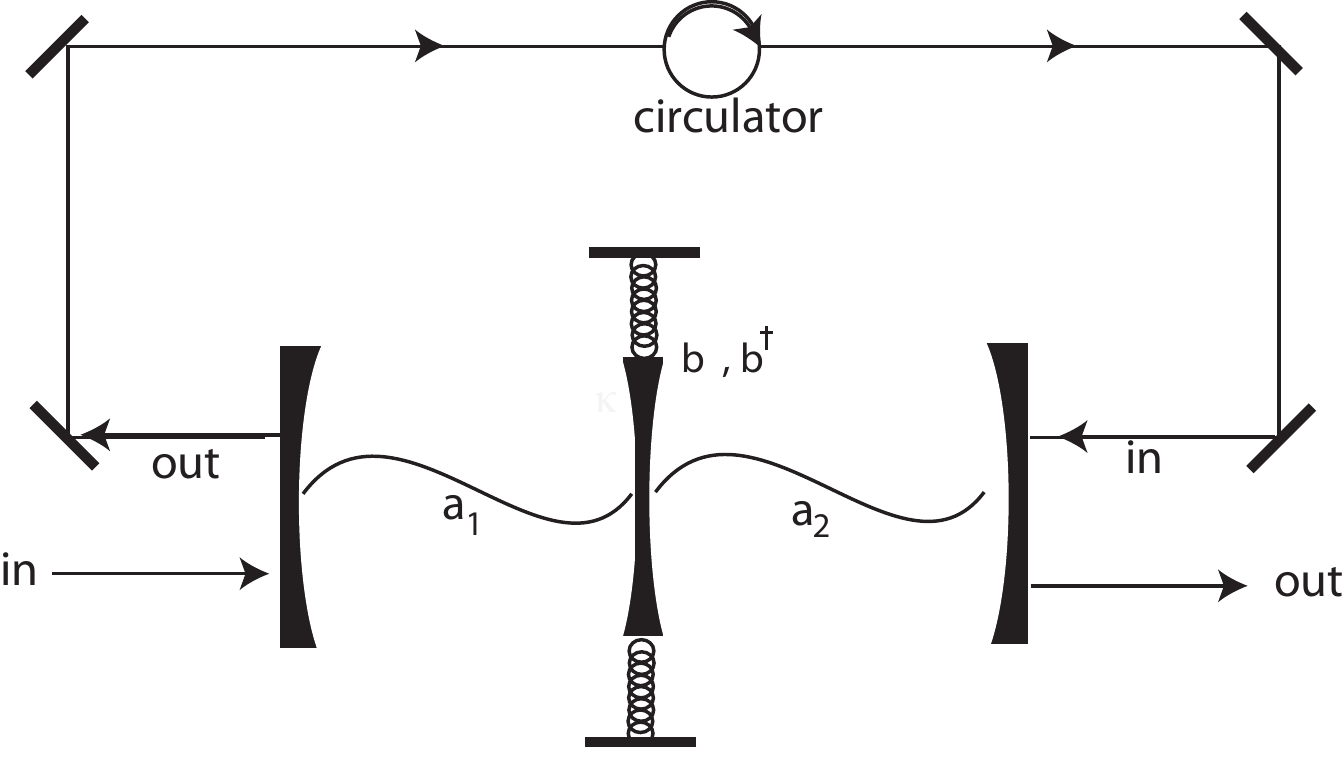}
\caption{Model 2: Implementing an all-optical feedback scheme in an optomechanical network: Two optical modes $\hat{a}_{1,2}$ are coupled reversibly via a common mechanical mode $\hat{b}$ forming an optomechanical network. An additional irreversible coupling is channelled with mode $a_{1}$ driving mode $a_{2}$ forming the feedback loop. The optomechanical interaction on either side of the mechanical element is denoted by a coupling strength $g_{j}$.}
\label{model_twosetp}
\end{figure} 
In the present section we shall discuss the effect of all-optical feedback  in a different setting  of coupled optical and mechanical modes. We consider a scenario where the two optical modes are coupled to a common mechanical mode with a bilinear interaction as shown in Fig.~\ref{model_twosetp}. The Hamiltonian describing the interaction between the three coupled bosonic modes thus takes the following form,
\begin{equation}\label{firsthamtn}
\tilde{H}=\sum_{j=1,2}\Delta_{j}\hat{a}_{j}^{\dagger} \hat{a}_{j} +\omega_{m} \hat{b}^{\dagger} \hat{b}+(g_{1}(\hat{a}_{1}^{\dagger}+ \hat{a}_{1} )+g_{2}(\hat{a}_{2}+\hat{a}_{2}^{\dagger} ))(\hat{b}+\hat{b}^{\dagger}),
\end{equation}
where $\Delta_{j}=\Omega_{j}-\omega_{L_{j}}$, where $\omega_{L_{j}}$ is the frequency of the external laser drive pumping  the $j$-th cavity whose resonance frequency is $\Omega_{j}$. The above Hamiltonian, for instance, can also describe the coherent coupling between three bilinearly coupled harmonic oscillators arranged in an open chain. In pursuit of main aim of this work  we consider a scenario where in addition to an indirect interaction between the two  optical  modes, the two optical cavity modes also interact  with a forward feed coupling.  Under the Born-Markov approximation, the open system dynamics of the coupled optomechanical cavities  then takes the following form,
\begin{eqnarray}\label{fulleqnthrtwo}
\frac{d \rho} {d t}=-i[\tilde{H},\rho]+\sqrt{\Gamma_{1} \Gamma_{2}} [\hat{a}_{1} \rho, \hat{a}_{2}^{\dagger} ]e ^{i (\omega_{L_{2}}-\omega_{L_{1}})t}
\\ \nonumber 
 +\sqrt{\Gamma_{1} \Gamma_{2}} [\hat{a}_{2} , \rho \hat{a}_{1}^{\dagger}]e ^{i (\omega_{L_{1}}-\omega_{L_{2}})t}+\sum_{j=1,2} \frac{\Gamma_{j}}{2}\mathcal L_{\hat{a}_{j}} \rho + \frac{\gamma_{1}}{2}(\bar{n}+1)\mathcal L_{\hat{b}} \rho+\frac{\gamma_{1}}{2}(\bar{n})\mathcal L_{\hat{b}^{\dagger}} \rho.
\end{eqnarray}
Working in the interaction picture of the bare frequency of the optical and the mechanical modes and further choosing the  physical parameters such that $\Delta_{1}=-\omega_{m}$ and  $\Delta_{2}=\omega_{m}$, then under the RWA the coupled system dynamics is governed by the following Master equation 
\begin{eqnarray}\label{fulleqnthr42}
\frac{d \rho} {d t}=-i g_{1}[\hat{a}_{1}^{\dagger}\hat{b}^{\dagger}+\hat{a}_{1}\hat{b},\rho]-i g_{2}[\hat{a}_{2}^{\dagger}\hat{b}+\hat{b}^{\dagger} \hat{a}_{2},\rho]
+\sqrt{\Gamma_{1} \Gamma_{2}} [\hat{a}_{1}  \rho, \hat{a}_{2}^{\dagger} ]
\nonumber\\
 +\sqrt{\Gamma_{1} \Gamma_{2}} [\hat{a}_{2} , \rho \hat{a}_{1}^{\dagger}]+\sum_{j=1,2} \frac{\Gamma_{j}}{2}\mathcal L_{\hat{a}_{j}} \rho + \frac{\gamma_{1}}{2}(\bar{n}+1)\mathcal L_{\hat{b}} \rho+\frac{\gamma_{1}}{2}(\bar{n})\mathcal L_{\hat{b}^{\dagger}} \rho.
\end{eqnarray}
where an explicit time independent form of the Master equation has been arrived at by choosing the physical parameters such that $\Omega_{1}=\Omega_{2}=(\omega_{L_{1}}+\omega_{L_{2}})/2$ and $\omega_{m}=(\omega_{L_{1}}-\omega_{L_{2}})/2$. Taking a similar approach to the previous section, it is possible to solve the Master equation \eref{fulleqnthr42} and the solution is outlined in the \ref{sec:appendixtwo}.

\begin{figure}[ht!]
     \begin{center}
    \subfigure[]{%
           \label{fig:secb}
           \includegraphics[width=0.45\textwidth]{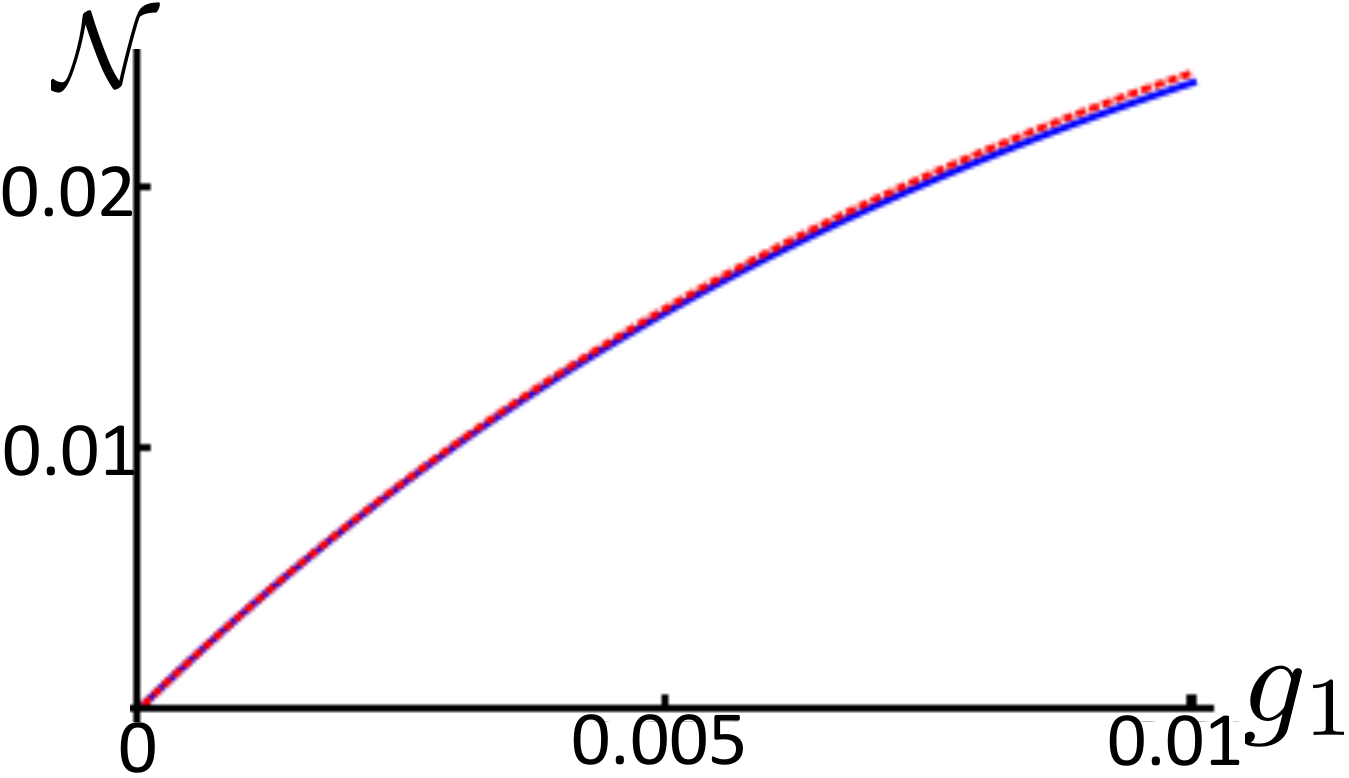}
        }
        \subfigure[]{%
           \label{fig:secc}
           \includegraphics[width=0.45\textwidth]{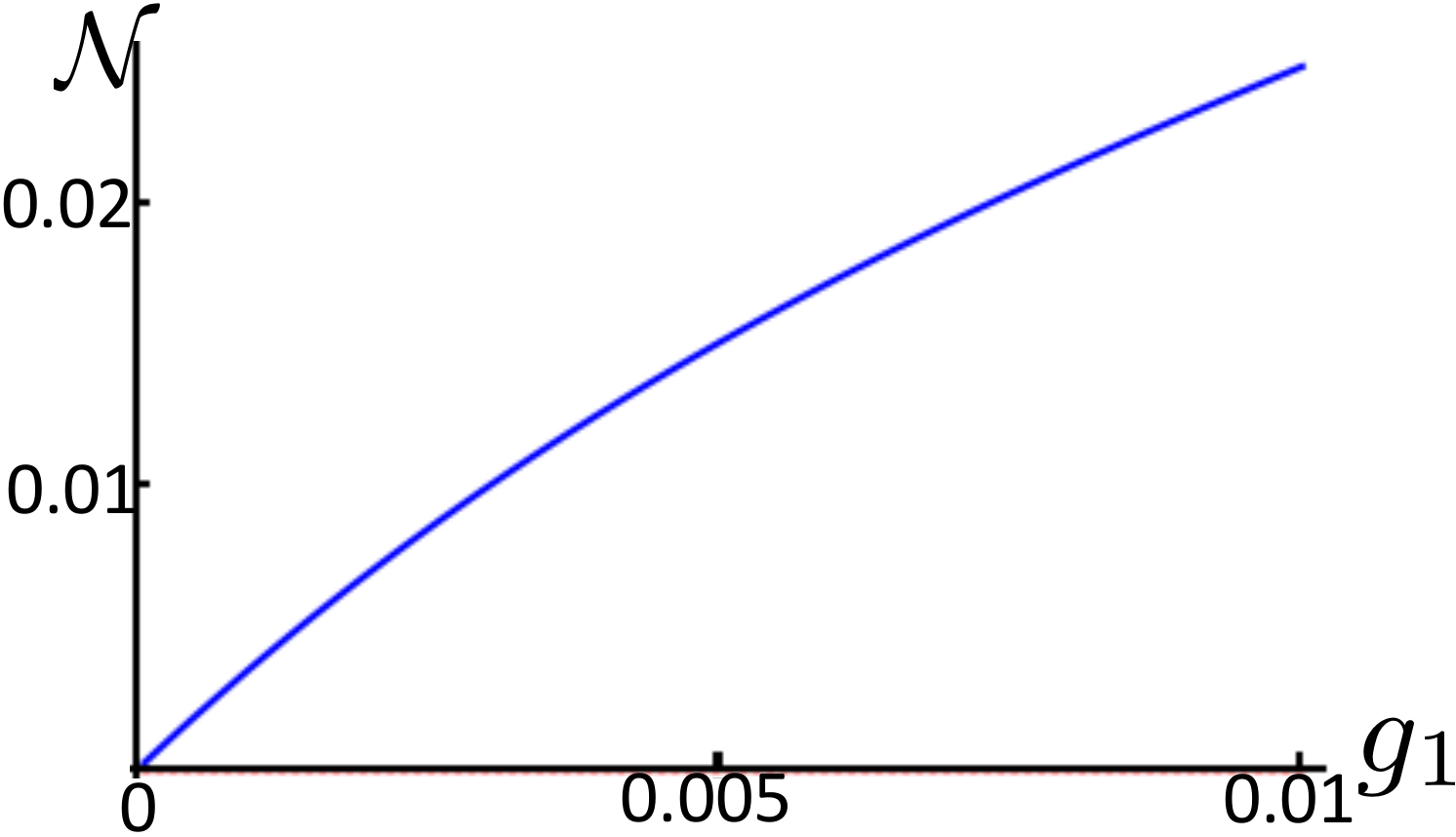}
        }
    \end{center}
    \caption{%
      (Color online)Steady state entanglement between the modes $(a)$ $\hat{a}_{1}, \hat{b}$ and $(b)$ $\hat{a}_{2}, \hat{b}$ plotted as a function of $g_{1}$  in presence (solid)and absence (dashed) of feedback. Physical parameters (in units of $\Gamma_{1}=\Gamma_{2}=\Gamma$): $g_{2}=0.05, \gamma_{1}=\gamma_{2}=10^{-2}, \bar{n}=0$.
     }%
   \label{fig:subfiguressecgeo}
\end{figure}

From the steady state solution of the Master equation \eref{fulleqnthr42} we compute the logarithmic negativity between the optical and mechanical modes. We plot the logarithmic negativity as an indicator of steady state entanglement between the optical and mechanical modes as a function of OM coupling strength $g_{1}$ in Fig~\ref{fig:subfiguressecgeo}. The entanglement between the modes $a_{1}$ and $b$ is plotted in Fig.~\ref{fig:secb}, while Fig.~\ref{fig:secc} shows the entanglement between the modes $a_{2}$ and $b$. Each figure compares how the steady state entanglement between the relevant mode varies in the presence as well as absence of all-optical feedback.  We note that feedback does not bring about qualitative  change in the steady state of optical and mechanical modes $a_{1},b$. On the other hand we observe that the presence of feedback generates entanglement between otherwise unentangled modes $a_2$ and $b$. The degree of entanglement between the modes $a_2$ and $a_1$ (nots shown here) remains relatively small both in the presence and absence of the all-optical feedback. 

Some more insight in understanding the role of all-optical feedback in generating quantum entanglement between modes  $a_{2}$ and $b$ can be obtained by writing Langevin's equations of motion for the modes $a_{2}$ and $b$.  In the absence of an all-optical feedback and  adiabatically eliminating the cavity mode $a_{1}$ results in 
\begin{eqnarray}
\frac{d}{dt}\hat{a}_{2}&=&-ig_{2}\hat{b}-\frac{\Gamma}{2}\hat{a}_{2}+\sqrt{\Gamma}\hat{a}_{in}(2,t)\\
\frac{d}{dt}\hat{b}&=&-ig_{2}\hat{a}_{2}-(\frac{\gamma}{2}-2\frac{g_{1}^{2}}{\Gamma})\hat{b}-\frac{2 i g_{1}}{\sqrt{\Gamma}}\hat{a}_{in}^{\dagger}(1,t)+\sqrt{\gamma}\hat{b}_{in}(1,t).
\end{eqnarray}
It is a straightforward exercise to arrive at  steady state expressions for the modes  $a_{2}$ and $b$. In the steady state one obtains
\begin{eqnarray*}\label{withnofedback}
\left(
\begin{array}{c}
 \hat{a}_{2}(\infty)\\
  \hat{b}(\infty)
\end{array}
\right)= \lim_{t \to \infty} \int_{0}^{\infty}&\left(
\begin{array}{cc}
 x_{1}(t') & x_{2}(t') \\
 x_{3}(t') & x_{4}(t')
\end{array}
\right). A(t-t')dt', 
\end{eqnarray*}
\begin{eqnarray*}
 A(t-t') =\left(
\begin{array}{c}
 \sqrt{\Gamma} \hat{a}_{in}(2,t-t') \\
(-2ig_{1}/\sqrt{\Gamma})\hat{a}_{in}^{\dagger}(1,t-t')+\sqrt{\gamma} \hat{b}_{in}(1,t-t')
\end{array}
\right)
\end{eqnarray*}
\begin{eqnarray}
\left(
\begin{array}{cc}
  x_{1}(t') & x_{2}(t') \\
 x_{3}(t') & x_{4}(t')
\end{array}
\right)&=&e^{Bt'}; B=\left(
\begin{array}{cc}
 -\Gamma/2 & -ig_{2} \\
-ig_{2} & -(\gamma/2-2g_{1}^{2}/\Gamma)
\end{array}
\right)
\end{eqnarray}
It is easy to verify that presence of completely uncorrelated noise operators $\hat{a}_{in}(1,t),\hat{a}_{in}(2,t)$ results in $\langle \hat{a}_{2}(\infty)\hat{b}(\infty)\rangle$=0. Thus in absence of all-optical feedback between the optical modes results in an uncorrelated steady state of  modes $a_{2}$ and $b$.
  
On the other hand, in the presence of all-optical feedback between the optical modes $a_{1}$ and $a_{2}$ adiabatically eliminating the cavity mode $a_{1}$ results in 
\begin{eqnarray}\label{withfedback}
\frac{d}{dt}\left(
\begin{array}{c}
 \hat{a}_{2} \\
 \hat{a}_{2}^{\dagger} \\
\hat{b} \\
\hat{b}^{\dagger}
\end{array}
\right)=\left(
\begin{array}{cccc}
 -\Gamma/2 & 0 & -ig_{2} & (2ig_{1}/\Gamma)\sqrt{\Gamma^{2}} \\
 0 &  -\Gamma/2 & -(2ig_{1}/\Gamma)\sqrt{\Gamma^{2}} & ig_{2}  \\
 -ig_{2}  & 0 &  -(\gamma/2 -2g_{1}^{2}/\Gamma)& 0 \\
 0 & ig_{2}  & 0 & -(\gamma/2 -2g_{1}^{2}/\Gamma)
\end{array}
\right).\left(
\begin{array}{c}
 \hat{a}_{2} \\
 \hat{a}_{2}^{\dagger} \\
\hat{b} \\
\hat{b}^{\dagger}
\end{array}
\right)\nonumber \\
+\left(
\begin{array}{c}
-\sqrt{\Gamma}\hat{a}_{in}(1,t) \\
-\sqrt{\Gamma}\hat{a}_{in}^{\dagger}(1,t) \\
\sqrt{\gamma}\hat{b}_{in}(1,t)-(2ig_{1}/\sqrt{\Gamma})\hat{a}_{in}^{\dagger}(1,t)  \\
\sqrt{\gamma}\hat{b}_{in}^{\dagger}(1,t)+(2ig_{1}/\sqrt{\Gamma})\hat{a}_{in}(1,t)
\end{array}
\right) ~~~
\end{eqnarray}
It is easy to obtain expressions for steady state correlations between modes $a_{2}$ and $b$. Assuming that the modes $a_{2}$ and $b$ are in contact with a zero-temperature reservoir one arrives at
\begin{eqnarray}
\langle \hat{a}_{2}(\infty)\hat{b}(\infty)\rangle=\int_{0}^{\infty}(\Gamma y_{1,1}y_{3,2}+2ig_{1}y_{1,1}y_{3,3}+\gamma y_{1,3}y_{3,4}-2ig_{1}y_{1,4}y_{3,2}+\frac{4g_{1}^{2}}{\Gamma}y_{1,4}y_{3,3})dt'~~~\nonumber\\
 \left(
\begin{array}{cccc}
y_{1,1}& y_{1,2}&y_{1,3} & y_{1,4} \\
y_{2,1}& y_{2,2}&y_{2,3} & y_{2,4} \\
y_{3,1}& y_{3,2}&y_{3,3} & y_{3,4} \\
y_{4,1}& y_{4,2}&y_{4,3} & y_{4,4} \\
\end{array}
\right)=e^{C t'};\nonumber \\
C
=\left(
\begin{array}{cccc}
 -\Gamma/2 & 0 & -ig_{2} & (2ig_{1}/\Gamma)\sqrt{\Gamma^{2}} \\
 0 &  -\Gamma/2 & -(2ig_{1}/\Gamma)\sqrt{\Gamma^{2}} & ig_{2}  \\
 -ig_{2}  & 0 &  -(\gamma/2 -2g_{1}^{2}/\Gamma)& 0 \\
 0 & ig_{2}  & 0 & -(\gamma/2 -2g_{1}^{2}/\Gamma)
\end{array}
\right).
\end{eqnarray}
Thus in the  presence of all-optical feedback between the optical modes $a_{1}$ and $a_{2}$,  we clearly have a non-vanishing steady state correlation function between the modes $a_{2}$ and $b$. This feature is corroborated in Fig.~\ref{fig:secc}.

\section{N-coupled optomechanical cavities}
\begin{figure}[ht!]
     \begin{center}
       {
            \includegraphics[width=0.8\textwidth]{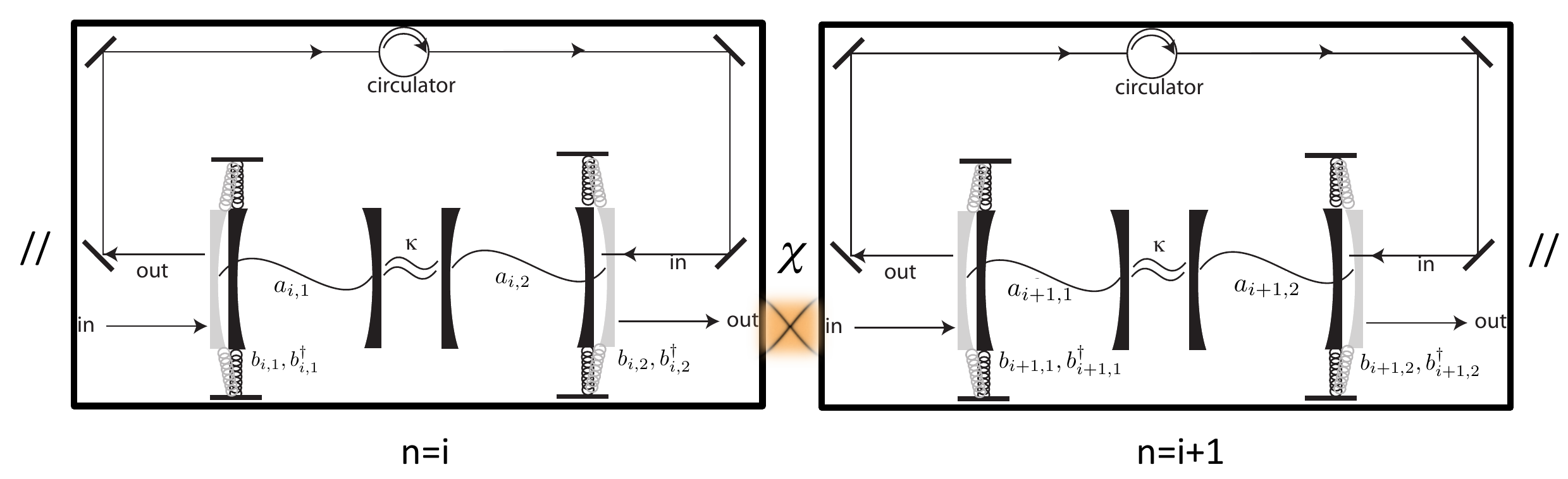}\\
        }
    \end{center}
    \caption{%
      (Color online) A schematic of a model of coupled OM array. Each box denotes an OM `port' which is coupled to its neighboring `ports' with coupling strength $\chi$.
     }\label{fig:chain}
\end{figure}

\begin{figure}[ht!]
     \begin{center}
       {
            \includegraphics[width=0.45\textwidth]{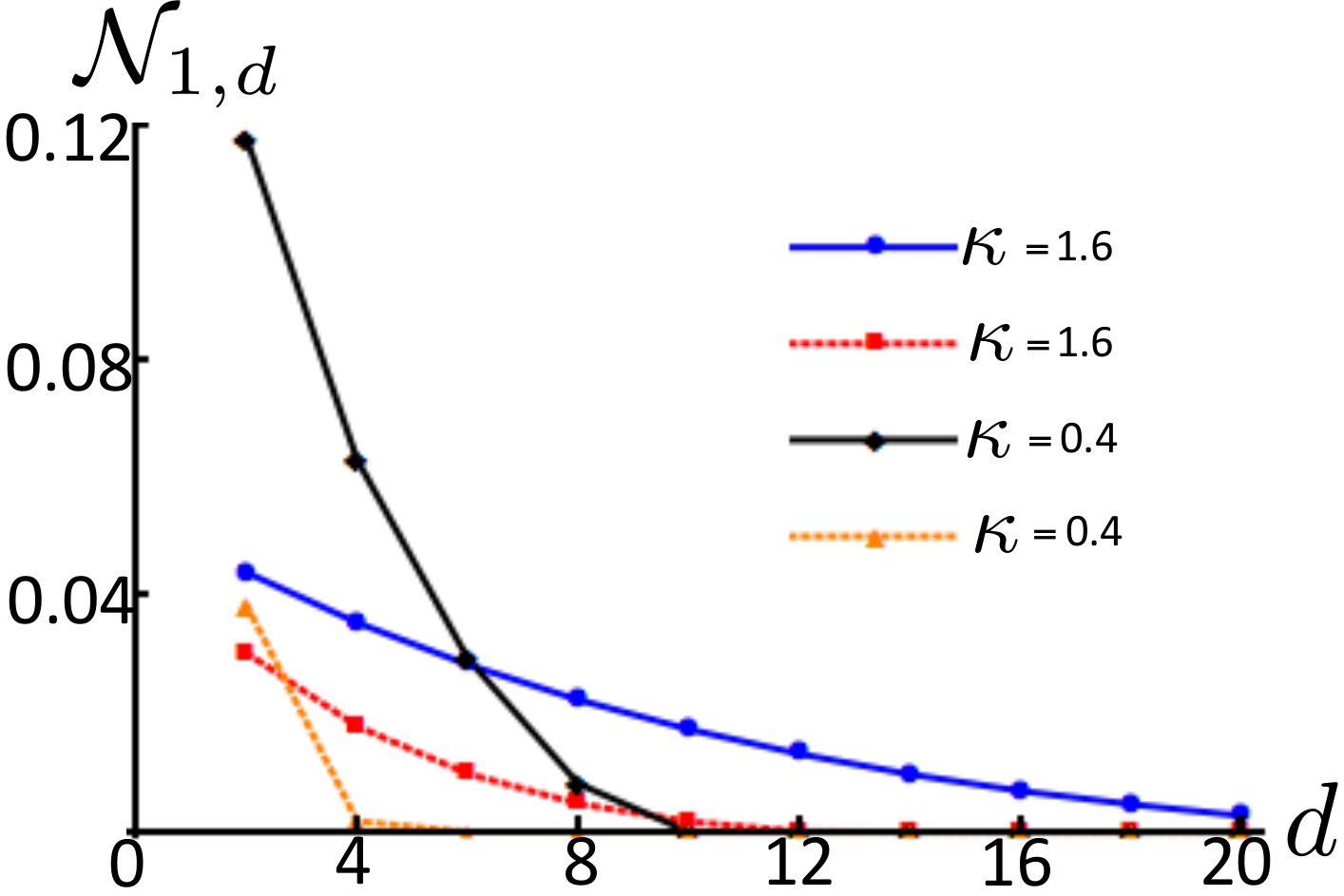}
        }
    \end{center}
    \caption{%
      (Color online) Steady state logarithmic negativity $\mathcal N_{1,d}$ plotted as a measure of entanglement between 1-st mechanical oscillator and  even-numbered oscillator in a chain of coupled OM `ports' in the presence (solid) and absence (dashed) of all optical feedback for different values of $\kappa$. Physical parameters (in units of $\Gamma_{1}=\Gamma_{2}=\Gamma$): $g_{1}=0.01, g_{2}=0.05, \gamma_{1}=\gamma_{2}=10^{-2}, \bar{n}=0, \chi=\kappa$ and number of ports $N=10$.
     }\label{fig:longchain}
\end{figure}

We now extend our analysis to study the influence of all-optical feedback on a large network of coupled OM cavities. We consider an array of N-OM ports. Each `port' itself comprises of a unit formed by two OM cavities coupled through an all-optical feedback, as described in section~\ref{model_1}. To invoke a genuine OM array with non-trivial quantum features in the steady state, we consider a scenario where there is a finite inter-mode coupling between different ports. In the absence of an inter-port coupling, quantum mechanical evolution of each individual port is described by the Master equation \eref{ME_model1}. 
Denoting $\rho_{n}$ for the state of $n^{\rm th}$ port in a network with N-coupled OM cavities, the joint state of the OM array  is thus simply given by $\rho =\otimes_{n=1}^N \rho_{n}$.
The optical and mechanical modes associated with an $i$-th `port' are labeled $a_{i,1},a_{i,2}$ and $b_{i,1},b_{i,2}$ respectively. The inter-port coupling between  $i$- th and $i$+1-th ports then takes the form 
\begin{equation}\label{intr_port}
V_{\rm int}=\chi (a^{\dagger}_{i,2}a_{i+1,1}+ a^{\dagger}_{i+1,1} a_{i,2}),
\end{equation}
where $\chi$ is the coupling strength. A schematic of such a model is shown in Fig.~\ref{fig:chain}. This sets the stage to understand the influence of all-optical feedback on a network of coupled OM cavities. We now numerically solve the Master equation for the network of coupled OM cavities with inter-port coupling of the form \eref{intr_port}, where each individual port is evolving under the Master equation \eref{ME_model1}. From the solution of the Master equation we compute the steady state entanglement between distant mechanical modes and the result is plotted in Fig.~\ref{fig:longchain}. Our simulations show that entanglement can only be generated between mechanical modes of different parity, i.e. $b_{i,1}$ and $b_{i+1,2}$.  This result arises because entanglement can only be generated between any two oscillators when one OM system driven on the blue sideband is coupled to the next OM system driven on the red sideband, and vice versa. However driving any two oscillators on the same sideband, does not result in a distribution or generation of entanglement, \cite{akram}. Therefore mechanical modes such as $b_{i,1}$ and $b_{i+1,1}$ remain separable for all parameters regardless of the absence or presence of the feedback loop.  

As illustrated in Fig.~\ref{fig:longchain}, we find once again that an all-optical feedback loop helps in generating stronger quantum entanglement between mechanical oscillators belonging to different ports. This observation follows from the results displayed in Fig.\ref{fig:subfigures} which show that the presence of feedback enhances the entanglement between the inter cavity mechanical modes in each port. Hence in the optomechanical array composed of N-ports, where each port has a similar topology to that discussed in section \ref{model_1}, we notice stronger entanglement between mechanical modes $b_{i,1}$ and $b_{i+1,2}$.  Another noteworthy feature of Fig.~\ref{fig:longchain} is that for smaller values of reversible coupling $\kappa$ between the optical modes, more entanglement is generated between distant mechanical modes. However, the spatial range of entanglement is short. On the other hand, for larger values of inter-optical coupling $\kappa$ ($\propto \chi$) spatial range of entanglement between the mechanical  modes grows. This, however, happens at the cost of reduced magnitude of entanglement between the mechanical modes. 

\section{Discussion and summary}
Optomechanics is rapidly emerging  as an  exciting branch of quantum technology where  it is possible to engineer unique hybrid quantum devices composed of seemingly incompatible physical systems. In the present work we have focussed on a regime where  each individual OM cavity is very strongly pumped and thus allowing us to linearize the inherently non-linear interaction between the optical and the mechanical modes. It should, however, be noted that there has also been interest in exploring the non-linear OM interaction by going beyond this linearization technique and thus allowing one to explore the non-Gaussian steady states of OM cavities \cite{Nunnenkamp, borke}. 

Expanding from previous work on distribution of entanglement in OM arrays, our scheme illustrates an all-optical feedback can be implemented to control the distributed entanglement in coupled OM arrays as well as generate otherwise non existent quantum correlations as in the mechanics assisted OM interaction scheme. It should also be noted that our control protocols apply equally well to hybrid optical-microwave schemes as in the recent experiments\cite{andrews}. Our scheme involving  an all-optical feedback has a clear advantage in generating spatially separated macroscopic entangled states. In the present work we have shown that  using a cascaded coupling between the optical modes, the degree of quantum correlations in an OM network can be further enhanced. As shown in Fig.~\ref{fig:subfigures} and Fig.~\ref{fig:subfigures1}, using an all-optical feedback results in generating stronger entanglement between distant mechanical modes and is more robust to thermal fluctuations of the mechanical resonators. The advantage  of using an all-optical feedback is clearly imprinted in the physical regime when the reversible  coupling between the optical modes is small. We believe this is not a handicap for our scheme and in fact provides more flexibility in choosing the separation between distant {\it nodes} of an OM network. For an alternate topology considered and as shown in  Fig.~\ref{fig:subfiguressecgeo}, we again find that an all-optical feedback may help in generating quantum correlations between distant optical and mechanical modes. We, however, do note that for  the range of parameters considered in Fig.~\ref{fig:subfiguressecgeo} steady state logarithmic negativity  is one order of magnitude smaller as compared to Fig.~\ref{fig:subfigures} and Fig.~\ref{fig:subfigures1}. This brings us closer to make a short remark on the physical meaning of non-zero value of logarithmic negativity and a possible way to enhance entanglement between the mechanical modes.

 As is well established now quantum entanglement is a vital requirement for the realization of  various tasks including quantum cryptography, quantum metrology, and quantum computing. However, entanglement is not very robust to environment induced decoherence and thus generally requires entanglement
distillation as an auxiliary tool to counteract the degradation of coherence.  Entanglement distillation refers to  extracting a small number of maximally entangled states from a larger ensemble of weakly entangled states \cite{distill}.  Logarithmic negativity, on the other hand,  provides an upper-bound on the amount of distillable entanglement \cite{eisertup}. Thus  one can make use of continuous variable entanglement distillation protocol to increase the entanglement of the shared state \cite{Opatrn}. It should be mentioned that for distilling Gaussian continuous variable states, non-Gaussian operations including photon addition/photon subtraction needs to be employed \cite{nogotheo}. An alternative approach is based on distilling Gaussian entanglement using quantum memory~\cite{adattali}

The ability to entangle a distributed array of mechanical elements will provide a path to new kinds of quantum enabled sensors. The Gaussian entanglement between harmonic oscillators that we have discussed in this paper is completely analogous to Gaussian entanglement for multi mode light fields. The enhanced metrology schemes that have been proposed in that case, for example~\cite{henning}, can carry over to the case of mechanical modes.

\begin{figure}[htpb]
  \centering
  \includegraphics[width=6.0in]{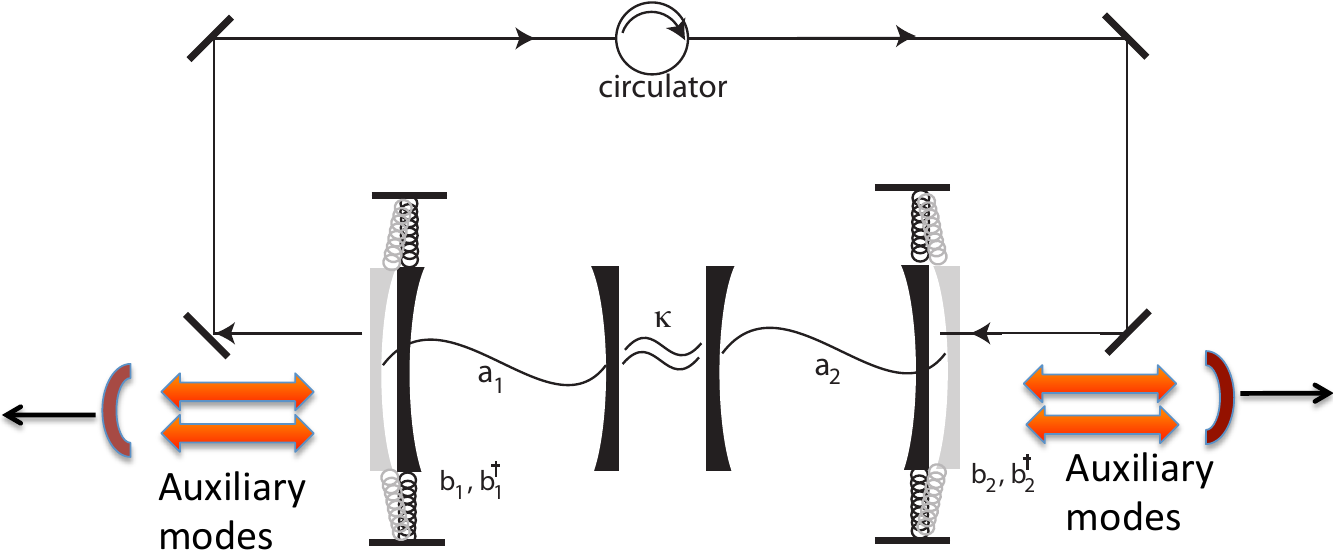}
  \caption{(Color online) A schematic of the scheme to detect quantum correlations between the mechanical modes of distant OM cavities.}
  \label{fig:detection}
\end{figure}

Since it is comparatively easier
to distill and detect  quantum correlations between optical modes,
as compared to directly detecting quantum entanglement
between mechanical modes, it has been suggested to swap the  the nonlocal correlations from the mechanical
modes back to the optical modes \cite{josh2,vitali, paternostro}.  As shown in Fig.~\ref{fig:detection}
this can, for instance, be implemented using two auxiliary
light modes, each initially prepared in classical uncorrelated
states. These auxiliary modes can be two modes
of distant cavities, and the geometry so arranged that
each entangled mirror couples independently to the two
modes. The non-local correlations may then be transferred
from the movable mirrors to the initially uncorrelated
auxiliary modes, which may eventually become entangled. Quantum correlated optical modes can then be eventually distilled. Thus, using standard homodyne measurement
techniques, the entire correlation matrix of the two optical 
auxiliary modes can be reconstructed. A presence
of non-zero quantum correlations between these optical
modes will be an indirect signature of non-zero quantum
correlations between the mechanical modes. The setup shown in Fig.~\ref{fig:detection} thus can also be used to ascertain the entanglement between the mechanical modes.

In our present work we have outlined a possibility of  generating remote macroscopic entanglement. Possibility of generating such macroscopic superpositions could shed light on understanding the mysteries of nature \cite{zure} and well be a useful resource in the construction of long-distance quantum communication networks \cite{Stannigel}. The model of OM network discussed in this work is general and we believe the ideas presented here could potentially be tested in number of physical settings. The parameters we have considered to describe the all-optical feedback scheme in this work do not place any stringent requirements on the degree of OM coupling strength nor the mechanical damping. Hence the ideas proposed in this paper would be suitable to a number of existing OM systems, \cite{Painter,Stamper-Kurn, Chan}. Comparing specific parameters from Chan et. al. \cite{Chan} where mechanical damping $\gamma=7.5kHz$ for cavity damping rate $\Gamma/2\pi=214MHz$ and OM coupling strength $g=1.1MHz$ fit in the regime of parameters we have considered in our calculations. A suitable system to realize our scheme  would be the implementation of cavity optomechanics using ultra-cold atoms as discussed in \cite{Stamper-Kurn}. Mechanics assisted coupling between the optical modes as required in the topology shown in  Fig.~\ref{model_twosetp} can be achieved using  collective vibrational degree of freedom of an ensemble of ultra-cold atoms \cite{Stamper-Kurn}. Another interesting test-bed for our results  could be provided by  optomechanical crystal (OMC) array which is a periodic structure that comprises both a photonic and a phononic crystal \cite{dechang}. Coherent coupling between the photons and phonons in the crystal provides the ingredients for the OM interaction discussed in this work. 

Other than creating massive quantum mechanical superposition of optical and mechanical modes, OM interaction provides other interesting avenues for implementing various tasks of quantum information processing. Storing and on-demand retrieval of quantum light, without compromising with its quantum character, has attracted wide attention both in the fields of  quantum optics and quantum information processing \cite{morgan,leon}. When it comes to implementing quantum light memory, electromagnetically induced transparency (EIT) is a much favored technique \cite{eitt}, where one can retain the large and highly desirable nonlinear optical properties associated with the resonant response of a material system. In analogy to the EIT in the atomic medium, optomechanically induced transparency (OMIT) has been experimentally demonstrated in \cite{kippenberg}. Moreover as illustrated in  \cite{dechang}, an optical waveguide can be coupled to an OMC array and quantum state of the light from the waveguide can be coherently transferred onto long-lived mechanical vibrations of the OMC array. Optomechanical   transducers  provides another promising application where coherent interaction between optics and mechanics can be exploited. As proposed in \cite{Stannigel}, an OM network can be employed to mediate interactions between distant nodes of a quantum network. Solid state qubits or electronic spin/charge  degrees of freedom could form nodes of the network. Coupling can be achieved between, $(a)$ the evanescent field of the microcavity and motion of the mechanical resonator, and $(b)$ also between  local nodes and vibrational degree of freedom of the mechanical resonator. Thus, indirect interaction mediated via mechanical resonator can then be set between optical mode and a distant node of the quantum network. 

To summarize we have analyzed the influence of an all-optical feedback loop on the steady state dynamics of OM arrays. We considered different topologies forming optomechanical networks and found that entanglement can be distributed over the chain of modes in the steady state and at finite temperatures. This opens up an interesting possibility to study spatially separated massive Schr\"odinger cat states.  In each case our results demonstrate that all-optical feedback can be used to enhance and thus {\it control} entanglement between inter cavity mechanical and optical modes. For the OM network resulting from two optical modes with a common mechanical element, we note that the presence of feedback in fact generates entanglement between the optical mode $a_2$ and mechanics $b$ which otherwise remain unentangled.  We have also extended the analysis to the case of an OM network comprised of N-OM ports. We have shown that an all-optical feedback also helps in distributing  stronger entanglement between inter cavity mechanical cavity modes. To conclude, we believe that the experimental  demonstration of  the quantum nature of macroscopic mechanical objects would help us to test long-standing questions about macroscopic quantum coherence and  long-lived mechanical states.

\section{Acknowledgments} {CJ acknowledges support from the ORS scheme and kind hospitality offered by the Centre for Engineered Quantum Systems, University of Queensland where this work was initiated. UA acknowledges support from the University of Queensland postdoctoral fellowship and grant. This work was supported by the Australian Research Council grant CE110001013.}

\appendix
\section{Solution of the Master equation for first optomechanical network}\label{sec:appendixone}
 To solve the Master equation \eref{ME_model1} we make a Gaussian ansatz for the quantum characteristic function of the following form  $\chi(\epsilon,\eta,x,y,{\it t}) =\exp\left[-z^{T}{\mbox{\bf A}}(t)\,z+i z^{T}h(t)\right]$, where ${\mbox{\bf A}}(t)$ is a time-dependent 8$\times$8 matrix and $h(t)$ is a 1$\times$8 time-dependent vector and  $z^{T}=(\epsilon,\epsilon^{*},\eta,\eta^{*},x,x^{*},y,y^{*})$.
The corresponding partial differential equation for $\chi(\epsilon,\eta,x,y,{\it t})$ then becomes 
\begin{equation}\label{chievnay}
\frac{\partial }{\partial t} \chi(\epsilon,\eta,x,y,{\it t})
=z^{T}  {\mbox{\bf N}} z\chi(\epsilon,\eta,x,y,{\it t}) +z^{T}  {\mbox{\bf M}} \nabla \chi(\epsilon,\eta,x,y,{\it t}),
\end{equation}
where
$\nabla= (\frac{\partial}{ \partial \epsilon},\frac{\partial}{ \partial \epsilon^{*}},\frac{\partial}{ \partial \eta},\frac{\partial}{ \partial \eta^{*}},\frac{\partial}{\partial x},\frac{\partial}{ \partial x^{*}},\frac{\partial}{ \partial y},\frac{\partial}{ \partial y^{*}}^{T})$
\begin{eqnarray}
{\mbox{\bf N}}&=&\left( \begin{array}{cccccccc}
    0&0&ig_{1}/2&0&0&0&0&0\\
    0&0&0&-ig_{1}/2&0&0&0&0\\
    ig_{1}/2&0&0&-\gamma \bar{n}/2&0&0&0&0\\
    0&-ig_{1}/2&-\gamma \bar{n}/2&0&0&0&0&0\\
    0&0&0&0&0&0&0&0\\
    0&0&0&0&0&0&0&0\\
    0&0&0&0&0&0&0&-\gamma \bar{n}/2\\
    0&0&0&0&0&0&-\gamma \bar{n}/2&0\\
  \end{array} \right )\\ \nonumber
{\mbox{\bf M}}&=&\left( \begin{array}{cccccccc}
    -\Gamma_{1}/2&0&0&-ig_{1}&i \kappa&0&0&0\\
    0&-\Gamma_{1}/2&ig_{1}&0&0&-i \kappa&0&0\\
    0&-ig_{1}&-\gamma /2&0&0&0&0&0\\
    ig_{1}&0&0&-\gamma /2&0&0&0&0\\
    i \kappa -\sqrt{\Gamma_{1} \Gamma_{2}}&0&0&0&-\Gamma_{2}/2&0&i g_{2}&0\\
    0&-i \kappa -\sqrt{\Gamma_{1} \Gamma_{2}}&0&0&0&-\Gamma_{2}/2&0&-ig_{2}\\
    0&0&0&0&ig_{2}&0&-\gamma/2&0\\
    0&0&0&0&0&-ig_{2}&0&-\gamma/2\\
  \end{array} \right ).
  \end{eqnarray}  
Using the Gaussian ansatz for the quantum characteristic function $\chi(\epsilon,\eta,x,y,{\it t})$,  it easily follows that,
\begin{equation}\label{diff}
\frac{\partial \chi }{ \partial t}=-z^{\rm T} \frac{d {\mbox{\bf A}}}{d t} z \chi+i z^{\rm T} \frac{d h}{d t} \chi, \nabla \chi =-{\rm 2} {\mbox{\bf A}} z \chi +i h \chi.
\end{equation}
Using \eref{diff}, the partial differential equation \eref{chievnay} for $\chi$ becomes,
\begin{equation}\label{difchmon}
-z^{\rm T} \frac{d {\mbox{\bf A}}}{d t} z\chi+i z^{\rm T} \frac{d h}{d t} \chi = z^{\rm T} {\mbox{\bf N}} z \chi-{\rm 2}z^{\rm T}{\mbox{\bf M}}{\mbox {\bf A}} z \chi +i z^{\rm T} {\mbox{\bf M}} h \chi.
\end{equation}
Recalling that $\mbox{\bf {A}}(t)$ is symmetric, the symmetric part of the \eref{difchmon} results in following two matrix differential equations,
   \begin{eqnarray}\label{nayitue}
-\frac{d {\mbox{\bf A}}(t)}{d t}=-{\mbox{\bf M}}{\mbox{\bf A}}-{\mbox{\bf A}} {\mbox{\bf M}}^{\rm T}+{\mbox{\bf N}} ; 
\frac{d h}{d t}= \mbox{\bf M} h
  \end{eqnarray}
 The above coupled matrix differential equations can now be numerically solved to get the time evolved quantum characteristic function for the coupled optomechanical cavities. 
For an initial Gaussian state evolving according to Master equation  \eref{nayitue},  it is sufficient to fully characterize the quantum correlations between various optical and mechanical modes in terms of their Wigner covariance matrix. The covariance matrix ${\mbox{\bf V}}$ is a $8 \times 8$ real symmetric matrix  $V_{i,j}=(\langle R_{i} R_{j}+R_{j} R_{i} \rangle )/2$  where $i,j \epsilon\{a,b\}$  and  $R^{T}=(\hat{q}_{a}, \hat{p}_{a},\hat{q}_{b},\hat{p}_{b})$. Here $\hat{q}_{i}$ and $\hat{p}_{i}$ are the position and momentum quadratures of the $i$th mode. From the expression of the quantum characteristic function it is straightforward to extract the Wigner covariance matrix as follows, $\langle {\hat{b}_{1}}^{\dagger m} {\hat{b}_{2}}^{n} \rangle= (\frac{\partial}{\partial \eta})^{m}(-\frac{\partial}{\partial y^{*}})^{n}\chi(\epsilon=0,x=0,{\it t})|_{\eta=0,y=0}$.


\section{Influence of feedback}\label{sec:appndxthr}
In this section we study in more detail how feedback affects the steady state correlations between the mechanical modes  $b_{1}$ and $b_{2}$. We focus our attention to a regime where each cavity mode has a large line width and can thus be adiabatically eliminated.  An equivalent way of describing the dynamics of coupled OM network array is to write the following  coupled Langevin equations,
\begin{eqnarray}
\frac{d}{dt}\hat{a}_{1}&=&-ig_{1}\hat{b}_{1}^{\dagger}-i\kappa\hat{a}_{2}-\frac{\Gamma}{2}\hat{a}_{1}+\sqrt{\Gamma}\hat{a}_{in}(1,t)\\
\frac{d}{dt}\hat{a}_{2}&=&-ig_{2}\hat{b}_{2}-i\kappa\hat{a}_{1}-\frac{\Gamma}{2}\hat{a}_{2}-\sqrt{\Gamma^{2}}\hat{a}_{1}+\sqrt{\Gamma}\hat{a}_{in}(1,t-\tau)\\
\frac{d}{dt}\hat{b}_{1}&=&-ig_{1}\hat{a}_{1}^{\dagger}-\frac{\gamma}{2}\hat{b}_{1}+\sqrt{\gamma}\hat{b}_{in}(1,t)\\
\frac{d}{dt}\hat{b}_{2}&=&-ig_{2}\hat{a}_{2}-\frac{\gamma}{2}\hat{b}_{2}+\sqrt{\gamma}\hat{b}_{in}(2,t),
\end{eqnarray}
where for simplifying the calculations we have assume that $\gamma_{1}=\gamma_{2}=\gamma$ and $\Gamma_{1}=\Gamma_{2}=\Gamma$. Neglecting the time delay between the source and the driven cavity $\tau$ and adiabatically eliminating both the cavity modes $\hat{a}_{1}$ and $\hat{a}_{2}$,  Langevin equations  for the mechanical modes become 
\begin{eqnarray*}
\frac{d}{dt}\hat{b}_{1}^{\dagger}=(-\frac{\gamma}{2}+\frac{\Gamma}{2}\frac{g_{1}^{2}}{\sigma^{*}})\hat{b}_{1}^{\dagger}+\sqrt{\gamma}\hat{b}_{ in}^{\dagger}(1,t)+(i\frac{g_{1}}{2 \sigma^{*}}\Gamma^{3/2}+g_{1}\frac{\kappa \sqrt{\Gamma}}{\sigma^{*}})\hat{a}_{in}-i \hat{g}_{1}\hat{g}_{2}\frac{\kappa}{\sigma^{*}}\hat{b}_{2}\\
\frac{d}{dt}\hat{b}_{2}=(-\frac{\gamma}{2}-\frac{\Gamma}{2}\frac{g_{2}^{2}}{\sigma^{*}})\hat{b}_{2}+\sqrt{\gamma}\hat{b}_{in}(2,t)+g_{1}g_{2}\frac{(\sqrt{\Gamma^{2}}+i\kappa)}{\sigma^{*}}\hat{b}_{1}^{\dagger}\nonumber \\
+(ig_{2}\sqrt{\Gamma}\frac{(\sqrt{\Gamma^{2}}+i\kappa)}{\sigma^{*}}-ig_{2}\frac{\Gamma^{3/2}}{2\sigma^{*}})\hat{a}_{in},
\end{eqnarray*}
where $\sigma=\Gamma^{2}/4+i\sqrt{\Gamma^{2}}\kappa+\kappa^{2}$.

After a lengthy but otherwise straightforward exercise it is possible to obtain the steady state correlators between the two mechanical modes. Assuming that both the cavity modes are in contact with a zero temperature reservoir then  the only non-zero correlator between the mechanical modes takes the form 
\begin{equation}\label{fedbckcorr}
\langle \hat{b}_{1} \hat{b}_{2} \rangle_{\rm feedback=0} =\frac{4 i \kappa g_{1}g_{2}}{\alpha^2}\gamma(\Gamma (g_{1}^2+g_{2}^2) \frac{2 \beta_{2}}{\beta_{1}}+\alpha)\frac{\beta_{2}}{\beta_{1}^{2}-4\beta_{2}^{2}}
\end{equation}
\begin{eqnarray}\label{nofedbckcorr}
\langle \hat{b}_{1} \hat{b}_{2} \rangle_{\rm feedback \ne 0} =\frac{4 (\sqrt{\Gamma^{2}}+i \kappa) g_{1}g_{2}}{\alpha^2-i16 \kappa \sqrt{\Gamma^{2}} g_{1}^{2}g_{2}^{2}}\gamma(\Gamma (g_{1}^2+g_{2}^2) \frac{2 \tilde{\beta}_{2}}{\tilde{\beta}_{1}}\nonumber \\
+\sqrt{\alpha^2-i16 \kappa \sqrt{\Gamma^{2}} g_{1}^{2}g_{2}^{2}})\frac{\tilde{\beta}_{2}}{\tilde{\beta}_{1}^{2}-4\tilde{\beta}_{2}^{2}},
\end{eqnarray}
where 
\begin{eqnarray}
\beta_{1}&=&\gamma+2\Gamma\frac{(g_{2}^{2}-g_{1}^{2})}{\Gamma^{2}+4\kappa^{2}}\\
\beta_{2}&=&\frac{\alpha}{\Gamma^{2}+4\kappa^{2}}\\
\tilde{\beta}_{1}&=&\gamma+2\Gamma\frac{(g_{2}^{2}-g_{1}^{2})}{\Gamma^{2}+4\kappa^{2}-i4\kappa\sqrt{\Gamma^{2}}}\\
\tilde{\beta}_{2}&=&\frac{\sqrt{\alpha^2-i16 \kappa \sqrt{\Gamma^{2}} g_{1}^{2}g_{2}^{2}}}{\Gamma^{2}+4\kappa^{2}-i4\kappa\sqrt{\Gamma^{2}}}\\
\alpha&=&\sqrt{\Gamma^{2}g_{1}^{4}+2(\Gamma^{2}+8\kappa^{2})g_{1}^{2}g_{2}^{2}+\Gamma^{2}g_{2}^{4}}
\end{eqnarray}

\section{Solution of the Master equation for second optomechanical network}\label{sec:appendixtwo}
  As done before,  to solve the Master equation \eref{fulleqnthr42} we  again convert it into a partial differential equation for the quantum characteristic function. Following the same steps as outlined in the previous section  we again define a normal-ordered characteristic function $\chi(\epsilon,\eta,x,{\it t})=\langle e^{\epsilon\hat{a}_{1}^{\dagger}}e^{-\epsilon^{*}\hat{a}_{1}}e^{\eta \hat{a}_{2}^{\dagger}}e^{-\eta^{*}\hat{a}_{2}}e^{x \hat{b}^{\dagger}}e^{-x^{*}\hat{b}}\rangle$. Now getting an expression for the time evolved quantum characteristic function reduces to solving the following coupled matrix differential equations
\begin{equation}\label{chievnay3two}
\frac{\partial }{\partial t} \chi(\epsilon,\eta,x,{\it t})
=z^{T}  {\mbox{\bf N}} z\chi(\epsilon,\eta,x,{\it t}) +z^{T}  {\mbox{\bf M}} \nabla \chi(\epsilon,\eta,x,{\it t}),
\end{equation}
where
$\nabla= (\frac{\partial}{ \partial \epsilon},\frac{\partial}{ \partial \epsilon^{*}},\frac{\partial}{\partial x},\frac{\partial}{ \partial x^{*}},\frac{\partial}{ \partial \eta},\frac{\partial}{ \partial \eta^{*}})^{T}$
\begin{eqnarray}
{\mbox{\bf N}}&=&\left( \begin{array}{cccccc}
    0&0&i g_{1}/2&0&0&0\\
    0&0&0&-i g_{1}/2&0&0\\
    i g_{1}/2&0&0&-\gamma_{1}\bar{n}/2&0&0\\
    0&-i g_{1}/2&-\gamma_{1}\bar{n}/2&0&0&0\\
   0&0&0&0&0&0\\
    0&0&0&0&0&0\\
  \end{array} \right )\\ \nonumber
{\mbox{\bf M}}&=&\left( \begin{array}{cccccc}
    -\Gamma_{1}/2&0&0&-ig_{1}&0&0\\
    0&-\Gamma_{1}/2&ig_{1}&0&0&0\\
    0&-ig_{1}&-\gamma_{1}/2&0&ig_{2}&0\\
  ig_{1}&0&0&-\gamma_{1}/2&0&-i g_{2}\\
     -\sqrt{\Gamma_{1} \Gamma_{2}}&0&ig_{2}&0&-\Gamma_{2}/2&0\\
    0&-\sqrt{\Gamma_{1} \Gamma_{2}}&0&-ig_{2}&0&-\Gamma_{2}/2\\
  \end{array} \right ).
  \end{eqnarray} 


\end{document}